%% file: main.tex
\documentclass[english,12pt]{article}\usepackage{lmodern} 

\usepackage[colorinlistoftodos,prependcaption,textsize=tiny,textwidth=0.8in]{todonotes}

\usepackage{etoolbox}
\newbool{arxiv}

\booltrue{arxiv}   

\usepackage{xr-hyper}

\notbool{arxiv}{
    \RequirePackage[colorlinks,
                    citecolor=blue,linkcolor=blue,urlcolor=blue,
                    pagebackref]{hyperref}
} {
    \usepackage{lineno}
    \usepackage{setspace}
    \onehalfspacing
    \makeatother
    \usepackage[unicode=true, pdfusetitle,
    bookmarks=true,bookmarksnumbered=false,bookmarksopen=false,
    breaklinks=true,pdfborder={0 0 0},backref=false]
    {hyperref}

    \usepackage[a4paper]{geometry}
    \geometry{verbose,tmargin=2.5cm,bmargin=2.5cm,lmargin=2.5cm,
              rmargin=2.5cm,columnsep=1cm}
}

\usepackage{refstyle}
\usepackage{varioref} 

\usepackage[T1]{fontenc}
\usepackage[latin9]{inputenc}
\usepackage{graphicx}
\usepackage{bbm}
\usepackage{subfigure}
\usepackage{fancyvrb}
\usepackage{bm}
\usepackage{lscape}
\usepackage{multirow}
\usepackage[authoryear]{natbib}
\usepackage{xcolor}
\usepackage{array,booktabs} 

\usepackage{xargs}[2008/03/08] 
\usepackage{mathrsfs} 

\usepackage[normalem]{ulem}

\usepackage{amssymb}
\usepackage{amsmath}
\usepackage{amsthm}

\usepackage{dcolumn}
\usepackage{lscape}
\usepackage{float}
\usepackage{mathtools}
\usepackage[titletoc,title]{appendix}
\usepackage{array}

\input{_math_defs}

\input{_reference_defs}
\input{_knitr_header}

\newbool{showlabels}
\boolfalse{showlabels}

\ifbool{showlabels}{
    \usepackage[outer]{showlabels}
    
    \usepackage{rotating}
    
}

\input{figures_knitr}
\DefineMacros{}

\begin{document}

\notbool{arxiv}{
\begin{frontmatter}
}

\title{An Automatic Finite-Sample Robustness Metric:
When Can Dropping a Little Data Make a Big Difference?}

\ifbool{arxiv}{
    \author{Tamara Broderick,
            Ryan Giordano\thanks{Equal contribution primary author}, and
            Rachael Meager\thanks{Equal contribution primary author}
        \footnote{
        Tamara Broderick and Ryan Giordano were supported in part by an
        Office of Naval Research Early Career Grant, an NSF CAREER Award, and an
        Army Research Office YIP Award.  We thank Avi Feller, Jesse Shapiro, Emily
        Oster, Michael Kremer, Peter Hull, Tetsuya Kaji, Heather Sarsons, Kirill
        Borusyak, Tin Danh Nguyen and the authors of all of our applications for
        their insightful comments and suggestions. All mistakes are our own.
        Corresponding Author: Rachael Meager, reachable at \texttt{r.meager@lse.ac.uk}.
        } 
    } 
    \maketitle
    \abstract{\input{abstract}}
}{ 
    \runtitle{When can Dropping a Little Data Make a Big Difference?}

    \begin{aug}
    \address[id=add1]{ \orgdiv{EECS}, \orgname{MIT}}
    \address[id=add2]{ \orgdiv{Economics}, \orgname{LSE}}
    \end{aug}

    \author[id=au1,addressref={add1}]
        {\fnms{T.}~\snm{Broderick}\ead[label=e3]
         {tbroderick@csail.mit.edu}}
    \author[id=au2,addressref={add1}]
        {\fnms{R.}~\snm{Giordano}\ead[label=e2]
         {rgiordan@mit.edu} \thanksref{t1}}
    \author[id=au3,addressref={add2}]
        {\fnms{R.}~\snm{Meager}\ead[label=e1]
         {r.meager@lse.ac.uk} \thanksref{t1}}

     \thankstext{t1}{Equal contribution first authors. }

    \support{
    Tamara Broderick and Ryan Giordano were supported in part by an
    Office of Naval Research Early Career Grant, an NSF CAREER Award, and an
    Army Research Office YIP Award.  We thank Avi Feller, Jesse Shapiro, Emily
    Oster, Michael Kremer, Peter Hull, Tetsuya Kaji, Heather Sarsons, Kirill
    Borusyak, Tin Danh Nguyen and the authors of all of our applications for
    their insightful comments and suggestions. All mistakes are our own.

    Corresponding Author: Rachael Meager, reachable at \texttt{r.meager@lse.ac.uk}.
    }

    \begin{abstract}
    \input{abstract}
    \end{abstract}


\end{frontmatter}
} 

\section{Introduction}
\seclabel{introduction}
\input{introduction}

\section{A proposed measure of sensitivity to dropping small data subsets}
\seclabel{metric}
    \seclabel{MIP}
    \input{MIP}

    \subsection{A Taylor series approximation to dropping data}
    \seclabel{taylor_series}
    \input{taylor_series}

    \subsection{A tractable approximation}
    \seclabel{AMIP}
    \input{AMIP}

    \subsection{Example functions of interest}
    \seclabel{function_examples}
    \input{function_examples}

    \subsection{A real-world OLS regression example}
    \seclabel{linear_regression}
    \input{introductory_regression_example}

\section{Underlying theory and interpretation} \seclabel{why}
    \input{theory_intro}

    \subsection{Theory and interpretation for Ordinary Least Squares}
    \seclabel{influence_function_ols}
    \input{influence_ols_example}

    \subsection{Theory and interpretation for general Z-estimators}
    \seclabel{influence_function}
    \input{influence_function}

    \subsection{Accuracy of the approximation}
    \seclabel{accuracy}
    \input{approximation_accuracy}

    \subsection{Related work}
    \seclabel{related_work}
    \input{related_work}

\section{Applied experiments} \seclabel{examples}

    \subsection{The Oregon Medicaid experiment}
    \seclabel{example_medicaid}
    \input{example_medicaid}

    \subsection{Cash transfers}
    \seclabel{example_transfers}
    \input{example_transfers}

    \subsection{Seven RCTs of microcredit: Linear regression analysis}
    \seclabel{example_microcredit_linear}
    \input{example_microcredit_linear}

    \subsection{Seven RCTs of microcredit: Bayesian hierarchical tailored mixture model}
    \seclabel{example_microcredit_hierarchical}
    \input{example_microcredit_hierarchical}

\section{Conclusion}\seclabel{conclusion}
\input{conclusion}

\clearpage
\newpage
\bibliography{robustness-lit}

\clearpage
\newpage
\begin{appendices}

\section{Detailed proofs}\applabel{proofs}

\input{appendix_proofs}

\section{The $\alpha$ dependence in \thmref{thetafun_accuracy} is tight}
\applabel{tight_bound}
\input{appendix_bound_tight}
\clearpage
\newpage


\end{appendices}

\end{document}

%% file: _math_defs.tex
\def\na{\texttt{NA}}

\DeclareMathOperator*{\argmax}{arg\,max}

\DeclarePairedDelimiter\abs{\lvert}{\rvert}
\DeclarePairedDelimiter\norm{\lVert}{\rVert}

\makeatletter
\let\oldabs\abs
\def\abs{\@ifstar{\oldabs}{\oldabs*}}

\def\norm{\@ifstar{\oldnorm}{\oldnorm*}}

\def\shape{\hat{\mathscr{T}}_\alpha}

\def\plim{\xrightarrow{p}}%
\def\dlim{\rightsquigarrow}%

\global\long\def\cop{C_{op}}%
\global\long\def\cgh{C_{gh}}%
\global\long\def\lh{L_{h}}%
\global\long\def\cij{C_{IJ}}%

\def\thetadom{\Omega_\theta}

\newcommand{\dee}{\mathrm{d}}

\def\infl{\psi}
\def\inflscale{\hat{\sigma}_{\infl}}
\def\inflscalelim{\sigma_{\infl}}
\def\mis#1{S_{#1}}
\def\mip#1{\Psi_{#1}}
\def\amis#1{\hat{S}_{#1}}
\def\amip#1{\hat{\Psi}_{#1}}
\def\loprop#1{\alpha^*_{#1}}
\def\aloprop#1{\hat{\alpha}^*_{#1}}
\def\thetafun{\phi}
\def\thetafunhat{\hat{\phi}}
\def\thetafunlin{\thetafun^{\mathrm{lin}}}%
\def\thetalin{\thetahat^{\mathrm{lin}}}%

\def\d{d} 
\def\p{p} 
\def\P{P} 
\def\w{\vec{w}}

\def\zP{0_{\P}}
\def\zPN{0_{\P \times N}}
\def\thetahat{\hat{\theta}}
\def\thetalim{\theta_{\infty}}
\def\onevec{\vec{1}}
\def\inflvec{\vec{\infl}}

\def\sumn{\sum_{n=1}^N}
\def\meann{\frac{1}{N}\sum_{n=1}^N}

\def\vnorm#1{\left\Vert #1\right\Vert }%
\def\at#1#2{\left.#1\right|_{#2}}%
\def\fracat#1#2#3{\at{\frac{#1}{#2}}{#3}}%
\def\expect#1{\mathbb{E}\left[#1\right]}%
\def\ind#1{\mathbb{I}\left(#1\right)}%

\def\maxover#1{\underset{#1}{\mathrm{max}}\,}

%% file: _reference_defs.tex
\theoremstyle{plain}
\newtheorem{lem}{Lemma}
\newtheorem{thm}{Theorem}

\newtheorem{assu}{Assumption}

\theoremstyle{definition}
\newtheorem{defn}{Definition}

\newcounter{point}[subsubsection]
\newenvironment{point}[1]%
{
\refstepcounter{point}
\textbf{(\alph{point}) #1.}
}%
{}

\newcommand{\pointlabel}[1]{\label{point:#1}}

\newcommand{\pointref}[1]{Point \ref{point:#1}}
\newcommand{\secpointref}[2]{\secref{#1}, paragraph \ref{point:#2}}

\newref{sec}{
    name=Section~, %
    Name=Section~,
    names=Sections~, %
    Names=Sections~
    }

\newref{app}{
    name=Appendix~, %
    Name=Appendix~
    }

\newref{eq}{
    name=Eq.~, %
    Name=Eq.~,
    names=Eqs.~, %
    Names=Eqs.~
    }

\newref{fig}{
    name=Figure~, %
    Name=Figure~,
    names=Figures~, %
    Names=Figures~
    }

\newref{table}{
    name=Table~, %
    Name=Table~,
    names=Tables~, %
    Names=Tables~
    }

\newref{def}{
    name=Definition~, %
    Name=Definition~,
    names=Definitions~, %
    Names=Definitions~
    }

\newref{assu}{
    name=Assumption~, %
    Name=Assumption~,
    names=Assumptions~, %
    Names=Assumptions~,
    }

\newref{cond}{
    name=Condition~, %
    Name=Condition~,
    names=Conditions~, %
    Names=Conditions~
    }

\newref{prop}{
    name=Proposition~, %
    Name=Proposition~,
    names=Propositions~, %
    Names=Propositions~
    }

\newref{lem}{
    name=Lemma~, %
    Name=Lemma~
    }

\newref{ex}{
    name=Example~, %
    Name=Example~,
    names=Examples~,
    Names=Examples
    }

\newref{cor}{
    name=Corollary~, %
    Name=Corollary~,
    names=Corollaries~,
    Names=Corollaries
    }

\newref{thm}{
    name=Theorem~, %
    Name=Theorem~
    }

\newref{proof}{
    name=Proof~, %
    Name=Proof~
    }

\newref{conj}{
    name=Conjecture~, %
    Name=Conjecture~
    }

\newref{algr}{
    name=Algorithm~, %
    Name=Algorithm~,
    names=Algorithms~, %
    Names=Algorithms~,
    }

%% file: _knitr_header.tex
\usepackage[]{graphicx}
\usepackage[]{color}
\makeatletter
\def\maxwidth{ %
  \ifdim\Gin@nat@width>\linewidth
    \linewidth
  \else
    \Gin@nat@width
  \fi
}
\makeatother

\definecolor{fgcolor}{rgb}{0.345, 0.345, 0.345}

\usepackage{framed}
\makeatletter
 {\par\unskip\endMakeFramed%
 \at@end@of@kframe}
\makeatother

\definecolor{shadecolor}{rgb}{.97, .97, .97}
\definecolor{messagecolor}{rgb}{0, 0, 0}
\definecolor{warningcolor}{rgb}{1, 0, 1}
\definecolor{errorcolor}{rgb}{1, 0, 0}
\newenvironment{knitrout}{}{} 

\usepackage{alltt}

%% file: figures_knitr.tex
\newcommand{\DefineMacros}{
\newcommand{\SimNumObs}{5,000}
\newcommand{\SimTrueTheta}{0.5}
\newcommand{\SimAccNumObs}{5,000}
\newcommand{\SimAccSigx}{2}
\newcommand{\SimAccSigeps}{1}
\newcommand{\SimAccPercentMax}{10}

}

\newcommand{\CashTransfersResultsTable}{

\begin{table}\scriptsize 
\begin{tabular}{|ccccc|}
\toprule
Study case & Original estimate  & Target change  & Refit estimate  & Observations dropped \\
\midrule
\midrule
& &  Sign change &  \textbf{-0.656 (3.745)} &  252 = 2.30\%\\
Poor, period 8  &  17.312 (4.576)* &  Significance change &  \textbf{7.284 (4.087)} &  83 = 0.76\%\\
& &  Significant sign change &  \textbf{-7.212 (3.443)*} &  464 = 4.24\%\\
\midrule
& &  Sign change &  \textbf{-1.377 (4.406)} &  345 = 3.58\%\\
Poor, period 9  &  27.924 (5.770)* &  Significance change &  \textbf{7.077 (4.555)} &  146 = 1.52\%\\
& &  Significant sign change &  \textbf{-8.951 (4.251)*} &  588 = 6.11\%\\
\midrule
& &  Sign change &  \textbf{-2.559 (3.541)} &  697 = 6.63\%\\
Poor, period 10  &  33.861 (4.468)* &  Significance change &  \textbf{4.806 (3.684)} &  435 = 4.14\%\\
& &  Significant sign change &  \textbf{-9.416 (3.296)*} &  986 = 9.37\%\\
\midrule
& &  Sign change &  \textbf{0.260 (6.410)} &  5 = 0.11\%\\
Non-poor, period 8  &  -5.444 (7.133) &  Significance change &  -12.845 (6.635) &  16 = 0.35\%\\
& &  Significant sign change &  9.670 (5.573) &  24 = 0.52\%\\
\midrule
& &  Sign change &  \textbf{-0.365 (7.542)} &  21 = 0.55\%\\
Non-poor, period 9  &  22.852 (10.000)* &  Significance change &  \textbf{16.506 (9.114)} &  3 = 0.08\%\\
& &  Significant sign change &  -11.733 (7.113) &  53 = 1.38\%\\
\midrule
& &  Sign change &  \textbf{-0.573 (6.750)} &  30 = 0.70\%\\
Non-poor, period 10  &  21.493 (9.405)* &  Significance change &  \textbf{16.262 (8.927)} &  3 = 0.07\%\\
& &  Significant sign change &  -10.845 (6.467) &  92 = 2.16\%\\
\midrule
\bottomrule
\end{tabular}
\caption{ Cash transfers results for various periods and treatment groups. The ``Refit estimate'' column shows the result of re-fitting the model removing the Approximate Most Influential Set. Stars indicate significance at the 5\% level.  Refits that achieved the desired change are bolded. }
\label{table:cash_transfers_re_run_table}
\end{table}

}

\newcommand{\OHIEResultsTable}{

\begin{table}\scriptsize 
\begin{tabular}{|ccccc|}
\toprule
Study case & Original estimate  & Target change  & Refit estimate  & Observations dropped \\
\midrule
\midrule
& &  Sign change &  \textbf{-0.006 (0.025)} &  275 = 1.18\%\\
Health genflip 12m  &  0.133 (0.026)* &  Significance change &  \textbf{0.044 (0.026)} &  162 = 0.69\%\\
& &  Significant sign change &  -0.043 (0.024) &  381 = 1.63\%\\
\midrule
& &  Sign change &  \textbf{-0.003 (0.015)} &  155 = 0.66\%\\
Health notpoor 12m  &  0.099 (0.018)* &  Significance change &  \textbf{0.027 (0.016)} &  100 = 0.43\%\\
& &  Significant sign change &  \textbf{-0.030 (0.015)*} &  219 = 0.94\%\\
\midrule
& &  Sign change &  \textbf{-0.006 (0.022)} &  197 = 0.84\%\\
Health change flip 12m  &  0.113 (0.023)* &  Significance change &  \textbf{0.039 (0.022)} &  106 = 0.45\%\\
& &  Significant sign change &  \textbf{-0.049 (0.022)*} &  291 = 1.24\%\\
\midrule
& &  Sign change &  \textbf{-0.023 (0.535)} &  73 = 0.33\%\\
Not bad days total 12m  &  1.317 (0.563)* &  Significance change &  \textbf{1.078 (0.558)} &  10 = 0.05\%\\
& &  Significant sign change &  -1.009 (0.521) &  144 = 0.66\%\\
\midrule
& &  Sign change &  \textbf{-0.040 (0.577)} &  87 = 0.41\%\\
Not bad days physical 12m  &  1.585 (0.606)* &  Significance change &  \textbf{1.131 (0.597)} &  20 = 0.09\%\\
& &  Significant sign change &  \textbf{-1.141 (0.566)*} &  164 = 0.77\%\\
\midrule
& &  Sign change &  \textbf{-0.062 (0.607)} &  123 = 0.57\%\\
Not bad days mental 12m  &  2.082 (0.640)* &  Significance change &  \textbf{1.171 (0.625)} &  42 = 0.19\%\\
& &  Significant sign change &  \textbf{-1.201 (0.594)*} &  212 = 0.98\%\\
\midrule
& &  Sign change &  \textbf{-0.005 (0.024)} &  123 = 0.53\%\\
Nodep Screen 12m  &  0.078 (0.025)* &  Significance change &  \textbf{0.046 (0.024)} &  42 = 0.18\%\\
& &  Significant sign change &  \textbf{-0.050 (0.023)*} &  220 = 0.95\%\\
\midrule
\bottomrule
\end{tabular}
\caption{ Medicaid profit results with IV for a range of outcome variables.  The ``Refit estimate'' column shows the result of re-fitting the model removing the Approximate Most Influential Set. Stars indicate significance at the 5\% level.  Refits that achieved the desired change are bolded. }
\label{table:ohie_profit_results_iv}
\end{table}

\begin{table}\scriptsize 
\begin{tabular}{|ccccc|}
\toprule
Study case & Original estimate  & Target change  & Refit estimate  & Observations dropped \\
\midrule
\midrule
& &  Sign change &  \textbf{-0.004 (0.008)} &  286 = 1.22\%\\
Health genflip 12m  &  0.039 (0.008)* &  Significance change &  \textbf{0.013 (0.008)} &  163 = 0.70\%\\
& &  Significant sign change &  \textbf{-0.021 (0.008)*} &  422 = 1.81\%\\
\midrule
& &  Sign change &  \textbf{-0.001 (0.005)} &  156 = 0.67\%\\
Health notpoor 12m  &  0.029 (0.005)* &  Significance change &  \textbf{0.008 (0.005)} &  101 = 0.43\%\\
& &  Significant sign change &  \textbf{-0.009 (0.004)*} &  224 = 0.96\%\\
\midrule
& &  Sign change &  \textbf{-0.002 (0.006)} &  198 = 0.85\%\\
Health change flip 12m  &  0.033 (0.007)* &  Significance change &  \textbf{0.011 (0.007)} &  106 = 0.45\%\\
& &  Significant sign change &  \textbf{-0.015 (0.006)*} &  292 = 1.25\%\\
\midrule
& &  Sign change &  \textbf{-0.013 (0.157)} &  74 = 0.34\%\\
Not bad days total 12m  &  0.381 (0.162)* &  Significance change &  \textbf{0.306 (0.161)} &  11 = 0.05\%\\
& &  Significant sign change &  \textbf{-0.309 (0.153)*} &  147 = 0.67\%\\
\midrule
& &  Sign change &  \textbf{-0.017 (0.169)} &  88 = 0.41\%\\
Not bad days physical 12m  &  0.459 (0.175)* &  Significance change &  \textbf{0.328 (0.172)} &  20 = 0.09\%\\
& &  Significant sign change &  \textbf{-0.344 (0.165)*} &  166 = 0.78\%\\
\midrule
& &  Sign change &  \textbf{-0.027 (0.178)} &  124 = 0.57\%\\
Not bad days mental 12m  &  0.603 (0.184)* &  Significance change &  \textbf{0.340 (0.181)} &  42 = 0.19\%\\
& &  Significant sign change &  \textbf{-0.381 (0.175)*} &  216 = 1.00\%\\
\midrule
& &  Sign change &  \textbf{-0.001 (0.007)} &  123 = 0.53\%\\
Nodep Screen 12m  &  0.023 (0.007)* &  Significance change &  \textbf{0.013 (0.007)} &  43 = 0.19\%\\
& &  Significant sign change &  \textbf{-0.015 (0.007)*} &  225 = 0.97\%\\
\midrule
\bottomrule
\end{tabular}
\caption{ Medicaid profit results with OLS for a range of outcome variables.  The ``Refit estimate'' column shows the result of re-fitting the model removing the Approximate Most Influential Set. Stars indicate significance at the 5\% level.  Refits that achieved the desired change are bolded. }
\label{table:ohie_profit_results_reg}
\end{table}

}

\newcommand{\MicrocreditProfitResultsTable}{

\begin{table}\scriptsize 
\begin{tabular}{|ccccc|}
\toprule
Study case & Original estimate  & Target change  & Refit estimate  & Observations dropped \\
\midrule
\midrule
& &  Sign change &  \textbf{-2.226 (15.628)} &  14 = 1.17\%\\
Bosnia  &  37.534 (19.780) &  Significance change &  \textbf{43.732 (18.889)*} &  1 = 0.08\%\\
& &  Significant sign change &  \textbf{-34.929 (14.323)*} &  40 = 3.35\%\\
\midrule
& &  Sign change &  \textbf{-0.053 (2.513)} &  1 = 0.03\%\\
Ethiopia  &  7.289 (7.893) &  Significance change &  \textbf{15.356 (7.763)*} &  45 = 1.45\%\\
& &  Significant sign change &  \textbf{-8.755 (1.852)*} &  66 = 2.12\%\\
\midrule
& &  Sign change &  \textbf{-0.501 (8.221)} &  6 = 0.09\%\\
India  &  16.722 (11.830) &  Significance change &  \textbf{22.895 (10.267)*} &  1 = 0.01\%\\
& &  Significant sign change &  \textbf{-16.638 (7.537)*} &  32 = 0.47\%\\
\midrule
& &  Sign change &  \textbf{0.398 (3.194)} &  1 = 0.01\%\\
Mexico  &  -4.549 (5.879) &  Significance change &  \textbf{-10.962 (5.565)*} &  14 = 0.08\%\\
& &  Significant sign change &  \textbf{7.030 (2.549)*} &  15 = 0.09\%\\
\midrule
& &  Sign change &  \textbf{0.021 (0.184)} &  16 = 1.66\%\\
Mongolia  &  -0.341 (0.223) &  Significance change &  \textbf{-0.436 (0.220)*} &  2 = 0.21\%\\
& &  Significant sign change &  \textbf{0.361 (0.147)*} &  38 = 3.95\%\\
\midrule
& &  Sign change &  \textbf{-0.569 (9.920)} &  11 = 0.20\%\\
Morocco  &  17.544 (11.401) &  Significance change &  \textbf{21.720 (11.003)*} &  2 = 0.04\%\\
& &  Significant sign change &  \textbf{-18.847 (9.007)*} &  30 = 0.55\%\\
\midrule
& &  Sign change &  \textbf{-4.014 (57.204)} &  9 = 0.81\%\\
Philippines  &  66.564 (78.127) &  Significance change &  \textbf{138.929 (66.880)*} &  4 = 0.36\%\\
& &  Significant sign change &  \textbf{-122.494 (49.409)*} &  58 = 5.21\%\\
\midrule
\bottomrule
\end{tabular}
\caption{ Microcredit regressions for the profit outcome. The ``Refit estimate'' column shows the result of re-fitting the model removing the Approximate Most Influential Set. Stars indicate significance at the 5\% level.  Refits that achieved the desired change are bolded. }
\label{table:mc_profit_results}
\end{table}

}

\newcommand{\MicrocreditTemptationResultsTable}{

\begin{table}\scriptsize 
\begin{tabular}{|ccccc|}
\toprule
Study case & Original estimate  & Target change  & Refit estimate  & Observations dropped \\
\midrule
\midrule
& &  Sign change &  \textbf{0.395 (2.135)} &  10 = 1.00\%\\
Bosnia  &  -5.803 (2.819)* &  Significance change &  \textbf{-4.870 (2.693)} &  1 = 0.10\%\\
& &  Significant sign change &  \textbf{5.130 (1.978)*} &  33 = 3.31\%\\
\midrule
& &  Sign change &  \textbf{0.035 (0.506)} &  41 = 0.60\%\\
India  &  -1.643 (0.576)* &  Significance change &  -1.051 (0.536)* &  8 = 0.12\%\\
& &  Significant sign change &  \textbf{1.059 (0.487)*} &  85 = 1.25\%\\
\midrule
& &  Sign change &  \textbf{0.000 (0.091)} &  12 = 0.07\%\\
Mexico  &  -0.082 (0.094) &  Significance change &  \textbf{-0.180 (0.091)*} &  14 = 0.09\%\\
& &  Significant sign change &  \textbf{0.176 (0.087)*} &  55 = 0.33\%\\
\midrule
& &  Sign change &  \textbf{-0.033 (0.973)} &  3 = 0.31\%\\
Mongolia  &  1.523 (2.103) &  Significance change &  \textbf{2.717 (1.027)*} &  10 = 1.04\%\\
& &  Significant sign change &  \textbf{-2.623 (0.689)*} &  45 = 4.68\%\\
\midrule
& &  Sign change &  \textbf{0.047 (0.669)} &  3 = 0.05\%\\
Morocco  &  -0.420 (0.723) &  Significance change &  \textbf{-1.351 (0.667)*} &  14 = 0.26\%\\
& &  Significant sign change &  \textbf{1.252 (0.602)*} &  23 = 0.42\%\\
\midrule
\bottomrule
\end{tabular}
\caption{ Microcredit regressions for the temptation outcome. The ``Refit estimate'' column shows the result of re-fitting the model removing the Approximate Most Influential Set. Stars indicate significance at the 5\% level.  Refits that achieved the desired change are bolded. }
\label{table:mc_temptation_results}
\end{table}

}

\newcommand{\MicrocreditMixtureResultsTable}{

\begin{table}\scriptsize 
\begin{tabular}{|ccccc|}
\toprule
Model parameter & Original estimate  & Target change  & Refit estimate  & Observations dropped \\
\midrule
\midrule
& &  Sign change &  \textbf{-0.042 (0.090)} &  31 = 0.09\%\\
$\tau_{-}$  &  0.102 (0.070) &  Significance change &  0.138 (0.071) &  11 = 0.03\%\\
& &  Significant sign change &  -0.204 (0.106) &  99 = 0.28\%\\
\midrule
& &  Sign change &  \textbf{-0.021 (0.046)} &  74 = 0.21\%\\
$\tau_{+}$  &  0.078 (0.033)* &  Significance change &  \textbf{0.062 (0.033)} &  9 = 0.03\%\\
& &  Significant sign change &  -0.100 (0.054) &  163 = 0.46\%\\
\midrule
\bottomrule
\end{tabular}
\caption{ Microcredit mixture results for a selected set of model parameters. Standard errors and ``significance'' are based on the estimated 95\% posterior credible intervals. The ``Refit estimate'' column shows the result of re-fitting the model removing the Approximate Most Influential Set. Stars indicate significance at the 5\% level.  Refits that achieved the desired change are bolded. }
\label{table:mcmix_re_run_table}
\end{table}

}

\newcommand{\MicrocreditMixtureSdResultsTable}{

\begin{table}\scriptsize 
\begin{tabular}{|cccccc|}
\toprule
Model parameter & Original estimate  & Change type  & Refit estimate  & Prediction  & Observations dropped \\
\midrule
\midrule
$\log \sigma_{\tau_{-}}$  &  -2.313 &  Drop 0.5\% to increase &  0.126 &  -0.811 &  177 = 0.50\%\\
& &  Drop 0.5\% to decrease &  -0.151 &  -4.066 &  177 = 0.50\%\\
\midrule
$\log \sigma_{\tau_{+}}$  &  -3.100 &  Drop 0.5\% to increase &  -1.095 &  -1.204 &  177 = 0.50\%\\
& &  Drop 0.5\% to decrease &  -1.598 &  -4.974 &  177 = 0.50\%\\
\midrule
\bottomrule
\end{tabular}
\caption{ Results for the log posterior standard deviation estimates of the effect size distribution in the microcredit mixture model.  Sign and significance are not meaningful for posterior standard deviations, so we drop 0.5\% of datapoints to attempt to produce large positive and negative changes.  The ``Refit estimate'' column shows the result of re-fitting the model removing Approximate Most Influential Set.  The ``Prediction'' column shows the predicted change under the same perturbation.   }
\label{table:mcmix_sd_re_run_table}
\end{table}

}

\newcommand{\SimCombNormalGraph}{

\begin{knitrout}
\definecolor{shadecolor}{rgb}{0.969, 0.969, 0.969}\color{fgcolor}\begin{figure}[!h]

{\centering \includegraphics[width=0.98\linewidth,height=0.524\linewidth]{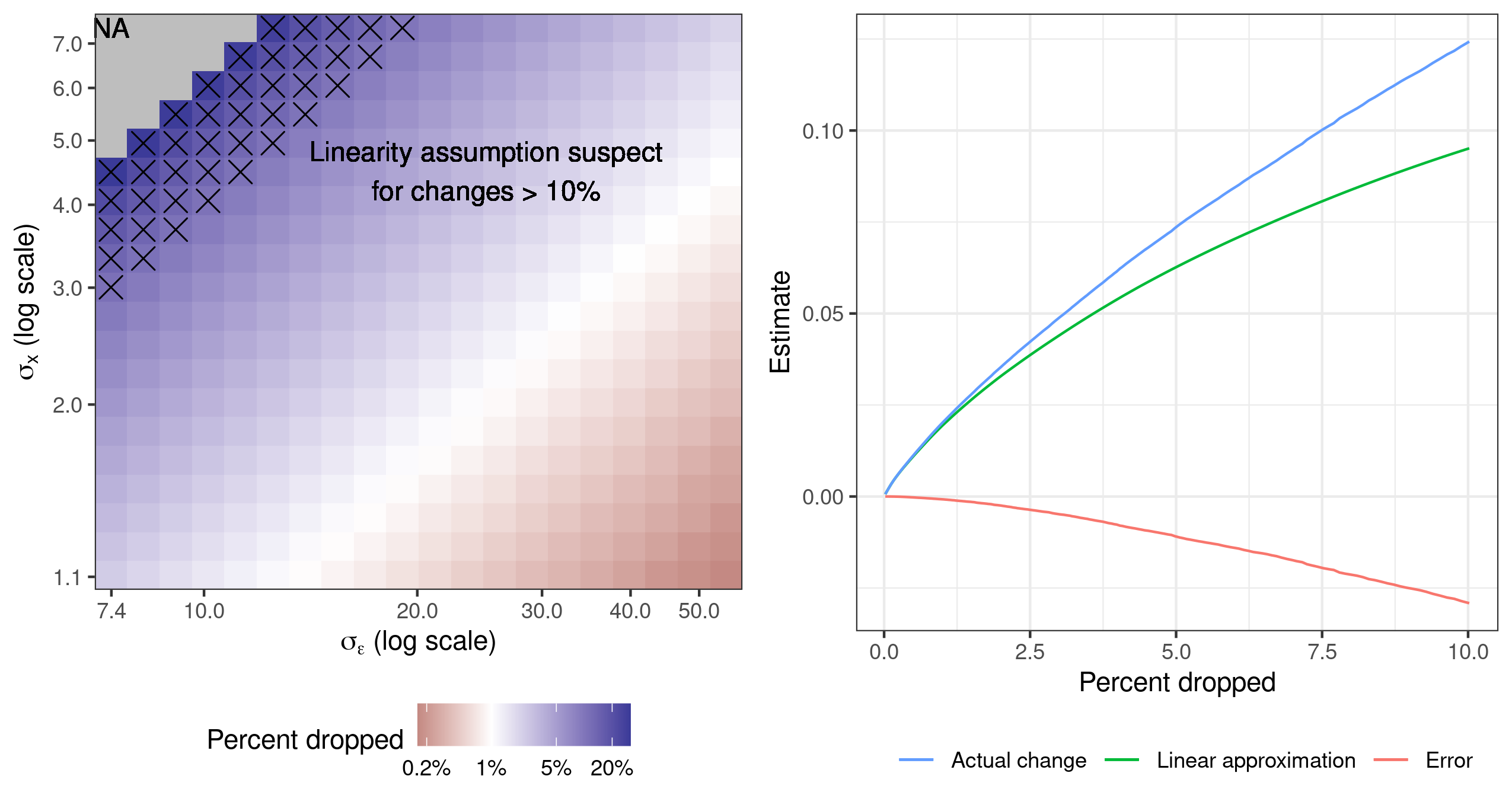} 

}

\caption{Simulation results for univariate linear regression with $N = \SimAccNumObs$ observations.  \textbf{Left panel:}  The approximate perturbation inducing proportion at differing values of $\sigma_x$ and $\sigma_\varepsilon$. Red colors indicate datasets whose sign can is predicted to change when dropping less than 1\% of datapoints. The grey areas indicate $\amip{\alpha} = \na$, a failure of the linear approximation to locate any way to change the sign.  \textbf{Right panel:}  The actual change, linear approximation to the change, and approximation error for $\sigma_x = \SimAccSigx$ and $\sigma_\varepsilon = \SimAccSigeps$.}\label{fig:sim-comb-normal}
\end{figure}

\end{knitrout}
}

%% file: abstract.tex
Study samples often differ from the target populations of inference and policy
decisions in non-random ways.  Researchers typically believe that such
departures from random sampling --- due to changes in the population over time and
space, or difficulties in sampling truly randomly --- are small, and their
corresponding impact on the inference should be small as well.  We might
therefore be concerned if the conclusions of our studies are excessively
sensitive to a very small proportion of our sample data.  We propose a method to
assess the sensitivity of applied econometric conclusions to the removal of a
small fraction of the sample.  Manually checking the influence of all possible
small subsets is computationally infeasible, so we use an approximation to find
the most influential subset.  Our metric, the ``Approximate Maximum Influence
Perturbation,'' is based on the classical influence function, and is
automatically computable for common methods including (but not limited to) OLS,
IV, MLE, GMM, and variational Bayes.   We provide finite-sample error bounds on
approximation performance.   At minimal extra cost, we provide an exact
finite-sample lower bound on sensitivity.   We find that sensitivity is driven
by a signal-to-noise ratio in the inference problem, is not reflected in
standard errors, does not disappear asymptotically, and is not due to
misspecification. While some empirical applications are robust, results of
several influential economics papers can be overturned by removing less than 1\%
of the sample.

%% file: introduction.tex
Ideally, policymakers will use economics research to inform decisions that
affect people's livelihoods, health, and well-being. Yet study samples may
differ from the target populations of these decisions in non-random ways,
perhaps because of practical challenges in obtaining truly random samples, or
because populations generally differ across time and place. When these
deviations from the ideal random sampling exercise are small, one might think
that the empirical conclusions would still hold in the populations affected by
policy. It therefore seems prudent to ask whether a small percentage of a
study's sample---or a handful of data points---has been instrumental in
determining its findings. In this paper we provide a finite-sample,
automatically-computable metric of how dropping a small amount of data can
change empirical conclusions. We show that certain empirical results from
high-profile studies in economics can be reversed by removing less than 1\% of
the sample even when standard errors are small, and we investigate why.

There are several reasons to care about whether empirical conclusions are
substantially influenced by small percentages of the finite sample. In practice,
even if we can sample from the population of direct interest, small percentages
of the data are missing; either surveyors and implementers cannot find these
individuals, or they refuse to answer our questions, or their answers get lost
or garbled during data processing. As this missingness cannot safely be assumed
random, researchers might care whether their substantive conclusions could
conceivably be overturned by a missing handful of data points. Similarly,
consumers of research who are concerned about potentially non-random errors in
sample construction at any stage of the analysis might be interested in this
metric as a measure of the exposure of a study's conclusions to this concern.
Conclusions that are highly influenced by a small handful of data points are
more exposed to adverse events or errors during data analysis, including
p-hacking, even if these errors are unintentional.

Even if researchers could construct a perfectly random sample from a given study
population, the target population for our policy decisions is almost always
different from the study population, if only because the world may change in the
time between the research and the decision. For this reason, social scientists
often aspire to uncover generalizable or ``externally valid'' truths about the
world and to make policy recommendations that would apply more broadly than to a
single study population.

In this paper, we propose to directly measure the extent to which a small
fraction of a data sample has influenced the central claims or conclusions of a
study. For a particular fraction $\alpha$ (e.g., $\alpha = 0.001$), we propose
to find the set of no more than $100 \alpha \%$ of all the observations that
effects the greatest change in an estimator when those observations are removed
from the sample, and to report this change. For example, suppose we were to
find a statistically-significant average increase in household consumption after
implementing some economic policy intervention. Further suppose that, by
dropping 0.1\% of the sample (often fewer than 10 data points), we instead find
a statistically-significant average \emph{decrease} in consumption. Then it
would be challenging to argue that there is strong evidence that this
intervention would yield consumption increases in even slightly different
populations.

To quantify this sensitivity, one could consider every possible $1-\alpha$
fraction of the data, and re-run the original analysis on all of these data
subsets. But this direct implementation is computationally
prohibitive.\footnote{Indeed, \citet{young2019consistency} finds it
computationally prohibitive to re-run their analysis when leaving out every
possible subset of two data points. To illustrate, consider an analysis that
takes 1 second to run; checking removal of every 4 data points from a data set
of size 400 would take over 33 years. See \secref{MIP} for more detail.} We
propose a fast approximation that works for common estimators---including
Generalized Methods of Moments (GMM), Ordinary Least Squares (OLS), Instrumental
Variables (IV), Maximum Likelihood Estimators (MLE), Variational Bayes (VB), and
all minimizers of smooth empirical loss (\secref{AMIP}).
Computation of the approximation is fast, automatable, and easy to use, and we
provide an \textsf{R} package on GitHub called
``zaminfluence.''\footnote{\url{https://github.com/rgiordan/zaminfluence}. The
name stands for ``Z-estimator approximate maximum influence.'' }

Our approximation is based on the classical ``influence function,'' which has
been used many times in the literature to assess sensitivity to dropping one or
a small number of datapoints (a discussion of related work can be found in
\secref{related_work} below).  However, prior work focused on outlier detection
and visual diagnostics and considered small numbers of removed datapoints.  In
contrast, we relate the effect of ablating a non-vanishing proportion of
datapoints to classical inference, with an interest in generalizing to unseen
populations rather than detection of gross outliers, and analyze the accuracy of
the empirical influence function as an approximation to leaving out a fixed
proportion of data.

Specifically, we show that our approximation performs well using a combination
of theoretical analyses, simulation studies, and applied examples. We
demonstrate theoretically that the approximation error is low when the
percentage of the sample removed is small (\secref{accuracy}). Moreover, for the
cost of a single additional data analysis, we can provide an exact lower bound
on the worst-case change in an analysis upon removing $100\alpha \%$ of the data
(\secref{exact_lower_bound}).  We check that our metric
detects combinations of data points that reverse empirical conclusions when
removed from real-life datasets (\secref{examples}). For example, in the Oregon
Medicaid study \citep{finkelstein2012oregon}, we can identify a subset
containing less than 1\% of the original data that controls the sign of the
effects of Medicaid on certain health outcomes. In the Mexico microcredit study
\citep{angelucci2015microcredit}, we find a single observation, out of 16,500,
that controls the sign of the ATE on household profit.

We investigate the source of this sensitivity when it arises, and we show that
it is not captured in conventional standard errors. We find that a result's
exposure to the influence of a small fraction of the sample need not reflect a
model misspecification problem nor the presence of gross outliers. Sensitivity
according to our metric can arise, even if the model is exactly correct and the
data set arbitrarily large, if there is a low \emph{signal-to-noise ratio}: that
is, if the strength of the claim (signal) is small relative to a quantity that
consistently estimates the standard deviation of the limiting distribution of
root-$N$ times the quantity of interest (\secref{why}). For example, in OLS this
``noise'' is large when we have a high ratio of residual variance to regressor
variance (\secref{influence_function_ols}). This noise can be large even when
standard errors are small, because it does not disappear as $N$ grows. This
result highlights the distinction between performing classical inference within
a hypothetical perfect random resampling experiment, and attempting to
generalize beyond the data to the world in which very small changes to the
population are occurring over space and time.

We examine several applications from empirical economics papers and find that
the sensitivity captured by our metric varies considerably across analyses in
practice. In many cases, the sign and significance of certain estimated
treatment effects can be reversed by dropping less than 1\% of the sample, even
when the t-statistics are very large and inference is very precise; see, e.g.,
the Oregon Medicaid RCT \citep{finkelstein2012oregon} in
\secref{example_medicaid}. In \secref{example_transfers}, we examine the
Progresa Cash Transfers RCT \citep{angelucci2009indirect} and show that trimming
outliers in the outcome data does not necessarily reduce sensitivity. In
\secref{example_microcredit_linear} we examine a simple two-parameter
linear regression on seven Microcredit RCTs \citep{meager2020aggregating} and,
in \secref{example_microcredit_hierarchical}, we examine a Bayesian hierarchical
analysis of the same data; these final two analyses show
that neither very simple nor relatively complex Bayesian models are immune to
sensitivity to dropping small fractions of the data.  However, not all analyses
we examine are non-robust.  Certain results across the applications we examine
are robust up to 5\% and even 10\% removal.

We recommend that researchers use our metric to complement standard errors and
other robustness checks. Our goal is not to supplant other sensitivity analyses,
but to provide an additional tool to be incorporated into a broader ecosystem of
systematic stability analysis in data science \citep{yu:2013:stability}. For
example, since our approximation is fundamentally local due to the Taylor
expansion, practitioners may also consider global sensitivity checks such as
those proposed by \citet{leamer1984global, leamer1985sensitivity,
sobol2001global,saltelli2004global}, or the conventional breakdown frontiers
approach of \citet{he1990tail, masten2020inference}.
Our method is also not a substitute for tailored robustness checks designed by
researchers to investigate specific concerns about sensitivity of results to
certain structures or assumptions. Applied researchers will always know more
than econometricians about which specific threats to their empirical strategies
are most worth investigating in order to solidify our trust in the results of
any given analysis.  And practitioners may well benefit from robustifying their
analysis \citep{mostellertukeydata, hansen2008robustness,chen2011sensitivity}
even if they pass our check.  Our metric is also complementary to classical
gross error robustness (which we take to include outlier detection and breakdown
point analyses) \citep{belsley:1980:regression,hampel1986robustbook}. In
particular, gross error sensitivity is designed to detect and accommodate
arbitrary adversarial perturbations to the population distribution.
We discuss similarities and differences between our work and other
robustness measures in detail in \secref{related_work}.

We do not recommend researchers discard results that are not robust to removal
of a small, highly-influential subset of data. While in certain cases such
sensitivity may be concerning for specific, contextually-determined reasons,
there is as yet no basis for doing so in general, as we have shown that such
sensitivity can arise even if the conventional inference is valid in the
strictest sense. However, we do suggest that researchers adjust their
interpretation of results which are sensitive to dropping a small fraction of
the data as being less generally applicable to somewhat differing populations,
and less robust to minor corruptions of their random sampling assumption. Much
as one would interpret statistically insignificant results as a failure to
detect an effect rather than positively detecting the absence of an effect,
sensitive results may indicate a failure to detect a transportable effect, but
not necessarily a failure of classical inference in itself. We do not yet
recommend any specific alterations to common inferential procedures based on our
metric, but we believe this direction is promising for future research.

%% file: MIP.tex


Suppose we observe $N$ data points $\d_{1}, \ldots, \d_{N}$. For instance, in a
regression problem, the $n$-th data point might consist of covariates $x_n$ and
response(s) $y_n$, with $d_n = (x_n,y_n)$. Consider a parameter $\theta \in
\mathbb{R}^{\P}$ of interest. Typically we estimate $\theta$ via some function
$\thetahat$ of our data. The central claim of an empirical economics paper is
typically focused on some attribute of $\theta$, such as the sign or
significance of a particular effect or quantity. A frequentist analyst might be
worried if removing some small fraction $\alpha$ of the data were to
\begin{itemize}
\item Change the sign of an effect.
\item Change the significance of an effect.
\item Generate a significant result of the opposite sign.
\end{itemize}
To capture these concerns, we define the following quantities:
\begin{defn} \deflabel{metrics}
Let the \emph{Maximum Influence Perturbation} be the largest possible change
induced in the quantity of interest by dropping no more than 100$\alpha$\% of
the data.

We will often be interested in the set that achieves the Maximum Influence
Perturbation, so we call it the \emph{Most Influential Set}.

And we will be interested in the minimum data proportion $\alpha \in [0,1]$
required to achieve a change of some size $\Delta$ in the quantity of interest,
so we call that $\alpha$ the \emph{Perturbation-Inducing Proportion}. We report
$\na$ if no such $\alpha$ exists.
\end{defn}

In general, to compute the Maximum Influence Perturbation for some $\alpha$, we
would need to enumerate every data subset that drops no more than 100$\alpha$\%
of the original data. And, for each such subset, we would need to re-run our
entire data analysis. If $m$ is the greatest integer smaller than 100$\alpha$,
then the number of such subsets is larger than $\binom{N}{m}$. For $N = 400$ and
$m=4$, $\binom{N}{m} = 1.05 * 10^9$. So computing the Maximum Influence
Perturbation in even this simple case requires re-running our data analysis over
1 billion times. If each data analysis took 1 second, computing the Maximum
Influence Perturbation would take over 33 years to compute. Indeed, the Maximum
Influence Perturbation, Most Influential Set, and Perturbation-Inducing
Proportion may all be computationally prohibitive even for relatively small
analyses.

To address this computational issue, we propose to instead use a (fast)
approximation to the Maximum Influence Perturbation, Most Influential Set, and
Perturbation-Inducing Proportion. We will see, for the cost of one additional
data analysis, our approximation can provide a lower bound on the exact Maximum
Influence Perturbation. More generally we provide theory and experiments to
support the quality of our approximation. We provide open-source
code\footnote{\url{https://github.com/rgiordano/zaminfluence}} and show that our
approximation is fully automatable in practice (\secref{AMIP}).

We articulate our approximation in \secref{taylor_series,AMIP} below.  First, in
\secref{taylor_series} to follow we derive a Taylor series approximation to the
act of leaving out datapoints.  Though this approximation is based on a
well-known first-order Taylor series approximation to the act of leaving out
datapoints, known as the {\em empirical influence function}
\citep{hampel1974influence,hampel1986robustbook}, we will assume no familiarity
with this work, deferring discussion of related literature to
\secref{influence_function,related_work}.  We then define our approximation to
data dropping in \secref{AMIP}, using the observation that the finding the
Maximum Influence Perturbation and its related quantities is trivial for the
Taylor series approximation.  We then conclude this section with some simple,
concrete examples of our approximation in
\secref{function_examples,linear_regression}.


%% file: taylor_series.tex
We begin by a deriving a Taylor series approximation to the act of dropping
data.  Though this approximation is well-known as the empirical influence
function (see \secref{influence_function} below for more details), we will
derive the approximation assuming no prior knowledge other than ordinary
multivariate calculus.

To form a Taylor series, we will naturally require certain aspects of our
estimator to be differentiable. We now summarize common assumptions under which
the Taylor expansion exists, and note that many common analyses satisfy these
assumptions---including, but not limited to, typical settings for OLS, IV, GMM,
MLE, and variational Bayes. Below, in \secref{accuracy}, we will state stricter
sufficient conditions that guarantee not only the existence but also the
finite-sample accuracy of our approximation.
\begin{assu}
$\thetahat$ is a \emph{Z-estimator}; that is, $\thetahat$ is the solution to the
following estimating equation,\footnote{Sometimes
\eqref{estimating_equation_no_weights} is associated with ``M-estimators'' that
optimize a smooth objective function, since such M-estimators typically take the
form of a Z-estimator which set the gradeint of the objective function to zero.
However, some Z-estimators, such as exactly identified IV regression or GMM, do
not optimize any particular empirical objective function, so the notion of
Z-estimator is in fact more general than that of an M-estimator.} where
$G(\cdot, \d_{n}): \mathbb{R}^{\P} \rightarrow \mathbb{R}^{\P}$ is a twice
continuously differentiable function and $\zP$ is the column vector of $P$
zeros.
\begin{align}\eqlabel{estimating_equation_no_weights}
\sumn G(\thetahat, \d_{n}) =  \zP .
\end{align}
\end{assu}
\begin{assu}
	$\thetafun: \mathbb{R}^{\P} \rightarrow \mathbb{R}$, which we interpret as a
	function that takes the full parameter $\theta$ and returns the quantity of
	interest from $\theta$, is continuously differentiable.\footnote{Below, we
	will allow for additional dependence in $\thetafun$ on data weights.}
\end{assu}
For instance, the function that picks out the $\p$-th effect from the vector
$\theta$, $\thetafun(\theta) = \theta_{\p}$, satisfies this assumption.

To form a Taylor series approximation to the act of leaving out datapoints, we
introduce a vector of data weights, $\w = (w_1, \ldots, w_N)$, where $w_n$ is
the weight for the $n$-th data point. We recover the original data set by giving
every data point a weight of 1: $\w = \onevec = (1, \ldots, 1)$. We can denote a
subset of the original data as follows: start with $\w = \onevec$; then, if the
data point indexed by $n$ is left out, set $w_n = 0$. We can collect weightings
corresponding to all data subsets that drop no more than 100$\alpha$\% of the
original data as follows:
\begin{align}\eqlabel{w_alpha_def}
	W_\alpha &:=
	\left\{ \w : \textrm{No more than }
 		   \lfloor \alpha N \rfloor \textrm{ elements of } \w \textrm{ are } 0
			\textrm{ and the rest are } 1 \right\}.
\end{align}
Our approximation will be to form a Taylor expansion of our quantity of interest
$\thetafun$ as a function of the weights, rather than recalculate $\thetafun$
for each data subset (i.e., for each reweighting).

To that end, we first reformulate our setup, now with the weights $\w$; note
that we recover the original problem (for the full data) above by setting
$\w=\onevec$ in what follows. Let $\thetahat(\w)$ be our parameter estimate at
the weighted data set described by $\w$. Namely, $\thetahat(\w)$ is the solution
to the weighted estimating equation
\begin{align} \eqlabel{estimating_equation_with_weights}
	\sumn w_n G(\thetahat(\w), \d_{n}) = \zP.
\end{align}
We allow that the quantity of interest $\thetafun$ may depend on $\w$ not only
via the estimator $\theta$, so we optionally write $\thetafun(\theta, \w)$
with $\thetafun(\cdot,\cdot): \mathbb{R}^{\P} \times \mathbb{R}^N \rightarrow
\mathbb{R}$.  Whenever we write $\thetafun(\cdot)$ as a function of a single
argument, we will implicitly mean $\thetafun(\cdot, \onevec)$.
We require that $\thetafun(\cdot,\cdot)$ be continuously differentiable in both
its arguments. For instance, we can use $\thetafun(\theta,\w) = \theta_{\p}$ to
pick out the $\p$-th component of $\theta$. Or, to consider questions of
statistical significance, we may choose $\thetafun(\theta,\w) = \theta_{\p} +
1.96 \sigma_{\p}(\theta,\w)$, where $\sigma_{\p}(\theta,\w)$ is an estimate of
the standard error depending smoothly on $\theta$ and $\w$; this example is our
motivation for allowing the more general $\w$ dependence in $\thetafun(\theta,
\w)$.

With this notation in hand, we can restate our original goal of computing
the Most Influential Set as solving
\begin{align} \eqlabel{mis_weight}
	\w^{**} &:=
	\argmax_{\w \in W_\alpha}
   		 \left( \thetafun(\thetahat(\w), \w) - \thetafunhat \right).
\end{align}
Here we focus on positive changes in $\thetafun$ since negative changes can be
found by reversing the sign of $\thetafun$ and using $-\thetafun$ instead. In
particular, the zero indices of $\w^{**}$ correspond to the Most Influential
Set: $\mis{\alpha} := \left\{n: \w^{**}_n = 0 \right\}$. And $\mip{\alpha} =
\thetafun(\w^{**}) - \thetafunhat$ is the Maximum Influence Perturbation. The
Perturbation Inducing Proportion is the smallest $\alpha$ that induces a change
of at least size $\Delta$: $\loprop{\Delta} := \inf\{ \alpha: \mip{\alpha} >
\Delta\}$.

%% file: AMIP.tex
Our approximation to the Maximum Influence Perturbation and its related
quantities, the Most Influential Set and Perturbation Inducing Proportion,
centers on a first-order Taylor expansion in $\w \mapsto
\thetafun(\thetahat(\w), \w)$ around $\w = \onevec$.  Let $\thetafunhat :=
\thetafun(\thetahat(\onevec), \onevec)$, the quantity of interest at the
original dataset.  Then:
\begin{align} \eqlabel{taylor_approx}
	\thetafun(\thetahat(\w), \w)
		&\approx \thetafunlin(\w)
		:= \thetafunhat +
            \sumn (w_n - 1) \infl_n,
    \textrm{ with } \infl_n :=
        \fracat{\partial \thetafun(\thetahat(\w), \w)}
               {\partial w_n}{\w = \onevec}.
\end{align}
We can in turn approximate the Most Influential Set as follows.  Let
$\infl_{(n)}$ denote the order statistics of $\infl_n$, i.e., the $\infl_n$
sorted from most negative to most positive.  Let $\ind{\cdot}$ denote the
indicator function taking value $0$ when the argument is false and $1$ when
true.  Then
\begin{align}
	\w^{**}  \approx
    \w^*
			:={}& \argmax_{\w \in W_\alpha}
   				 \left( \thetafunlin(\w) - \thetafunhat \right)
			= \argmax_{\w \in W_\alpha} \sum_{n: \, w_n = 0} \left(- \infl_n\right)
            \Rightarrow \nonumber
\\
\thetafunlin(\w^*) - \thetafunhat
    ={}& -\sum_{n=1}^{\lfloor \alpha N \rfloor} \infl_{(n)}
        \ind{\infl_{(n)} < 0}. \eqlabel{w_approx_opt}
\end{align}
To compute $\w^*$ (analogous to the $\w^{**}$ that determines the exact Most
Influential Set), we compute $\infl_n$ for each $n$. Then we choose $\w^*$ to
have entries equal to zero at the $\lfloor \alpha N \rfloor$ indices $n$ where
$\infl_n$ is most negative (and to have entries equal to one elsewhere).
Analogous to the Perturbation Inducing Proportion, we can find the minimum data
proportion $\alpha$ required to achieve a change of some size $\Delta$: i.e.,
such that $\thetafunlin(\w^*) - \thetafunhat > \Delta$. In particular, we
iteratively remove the most negative $\infl_n$ (and the index $n$) until the
$\Delta$ change is achieved; if the number of removed points is $M$, the
proportion we report is $\alpha = M/N$. Recall that finding the exact Maximum
Influence Perturbation, Most Influential Set, and Perturbation-Inducing
Proportion required running a data analysis more than $\binom{M}{\lfloor \alpha
N \rfloor}$ times. By contrast, our approximation requires running just the
single original data analysis, $N$ additional fast calculations to compute each
$\infl_n$, and finally a sort on the $\infl_n$ values.

We define our approximate quantities, as detailed immediately above, as follows.
\begin{defn} \deflabel{approx_metrics}
The \emph{Approximate Most Influential Set} is the set $\amis{\alpha}$ of at
most 100$\alpha$\% data indices that, when left out, induce the biggest
approximate change $\thetafunlin(\w) - \thetafunhat$; i.e., it is the set of
data indices left out by $\w^*$: $\amis{\alpha} := \left\{n: \w^{*}_n = 0
\right\}$.

The \emph{Approximate Maximum Influence Perturbation (AMIP)} $\amip{\alpha}$ is
the approximate change observed at $\w^*$: $\amip{\alpha} :=
\thetafunlin(\w^{*}) - \thetafunhat$.

The \emph{Approximate Perturbation Inducing Proportion} $\aloprop{\Delta}$ is
the smallest $\alpha$ needed to cause the approximate change $\thetafunlin(\w) -
\thetafunhat$ to be greater than $\Delta$. That is, $\aloprop{\Delta} := \inf\{
\alpha: \amip{\alpha} > \Delta\}$. We report $\na$ if no $\alpha \in [0,1]$ can
effect this change.
\end{defn}

Below, we will sometimes emphasize that the AMIP is a sensitivity and refer to
it as the \emph{AMIP sensitivity}. We will say that an analysis is
\emph{AMIP-non-robust} if, for a particular $\alpha$ of interest, the AMIP is
large enough to change the substantive conclusions of the analysis.  Conversely,
if the AMIP is not large enough, we say an analysis is \emph{AMIP-robust}.
And we generically use the AMIP acronym to describe our
methodology even when calculating the Approximate Most Influential Set or
Approximate Perturbation Inducing Proportion.

\subsubsection{An exact lower bound on the Maximum Influence Perturbation}
\seclabel{exact_lower_bound}

For any problem where performing estimation a second time is not prohibitively
costly, we can re-run our analysis without the data points in the Approximate
Most Influential Set and thereby provide a lower bound on the exact Maximum
Influence Perturbation.

Formally, let $\w^{**}$ be the weight vector for the exact Most Influential Set,
and let $\w^*$ be the weight vector for the Approximate Most Influential Set
$\amis{\alpha}$. We run the estimation procedure an extra time to recover
$\thetafun(\thetahat(\w^{*}), \w^{*})$. Then, by definition,
\begin{align*}
	\mip{\alpha} = \thetafun(\thetahat(\w^{**}), \w^{**}) - \thetafunhat
		&=
                    \maxover{\w \in W_\alpha}
                    \left(\thetafun(\thetahat(\w), \w) - \thetafunhat \right)
                    \ge
                    \thetafun(\thetahat(\w^{*}), \w^{*}) - \thetafunhat.
\end{align*}
Since $\thetafun(\thetahat(\w^{*}), \w^{*}) - \thetafunhat$ is a lower bound for
$\mip{\alpha}$, we can use the Approximate Most Influential Set to
conclusively demonstrate non-robustness. Of course, this lower bound holds for
{\em any} weight vector and will be most useful if the Approximate Maximum
Influence Perturbation is close to the exact Maximum Influence Perturbation. In
\secref{accuracy} below, we establish the accuracy of the approximation for small
$\alpha$ under mild regularity conditions.

\subsubsection{Computing the influence scores}

To finish describing our approximation, it remains to detail how to compute
$\infl_n = \fracat{\partial \thetafun(\thetahat(\w), \w)}{\partial
w_n}{\w=\onevec}$ from \eqref{taylor_approx}. We will refer to the quantity
$\fracat{\partial \thetafun(\thetahat(\w), \w)}{\partial w_n}{\w}$ as the
\emph{influence score} of data point $n$ for $\thetafun$ at $\w$ since, as we
discuss in \secref{influence_function} below, it is the \emph{empirical
influence function} evaluated at the datapoint $\d_{n}$. To compute the
influence score, we first apply the chain rule:
\begin{align} \eqlabel{chain_rule_influence_score}
	\fracat{\partial \thetafun(\thetahat(\w), \w)}{\partial w_n}
           {\thetahat(\w), \w}
		&=  \fracat{\partial \thetafun(\theta, \w)}{\partial \theta^T}{\thetahat(\w), \w}
   			 \fracat{\partial \thetahat(\w)}{\partial w_n}{\w} +
  			\fracat{\partial \thetafun(\theta, \w)}{\partial w_n}{\thetahat(\w), \w}.
\end{align}
The derivatives of $\thetafun(\cdot,\cdot)$ can be calculated using automatic
differentiation software
\citep{baydin2017automatic,tensorflow:2015:whitepaper,jax:2018:github,pytorch:2019:lots}.
And once we have $\thetahat(\onevec)$ from running the original data analysis,
we can evaluate these derivatives at $\w = \onevec$: e.g., $\fracat{\partial
\thetafun(\theta, \w)}{\partial \theta^T}{\thetahat(\onevec), \w=\onevec}$.

The term $\fracat{\partial \thetahat(\w)}{\partial w_n}{\w = \onevec}$ requires
slightly more work since $\thetahat(\w)$ is defined implicitly. We follow
standard arguments from the statistics and mathematics literatures
\citep{krantz2012implicit, hampel1974influence} to show how to calculate it
below.

Start by considering the more general setting where $\thetahat(\w)$ is the
solution to the equation $\gamma(\thetahat(\w), \w) =  \zP $. We assume
$\gamma(\cdot, \w)$ is continuously differentiable with full-rank Jacobian
matrix; then the derivative $\fracat{\partial \thetahat(\w)}{\partial w_n}{\w}$
exists by the implicit function theorem \citep[Theorem
3.3.1]{krantz2012implicit}. We can thus use the chain rule and solve for
$\fracat{\partial \thetahat(\w)}{\partial w_n}{\w}$; in what follows, $\zPN$ is
the $\P \times N$ matrix of zeros.
\begin{align}
	\zPN &= \fracat{\dee \gamma(\thetahat(\w), \w)}{\dee \w^T}{\w}
		= \fracat{\partial \gamma(\theta, \w)}{\partial \theta^T}{\thetahat(\w), \w}
\fracat{\dee \thetahat(\w)}{\dee \w^{T}}{\w} +
\fracat{\partial \gamma(\theta, \w)}{\partial \w^T}{\thetahat(\w), \w} \\
\Rightarrow
	\eqlabel{dtheta_dw_general}
	\fracat{\dee \thetahat(\w)}{\dee \w^{T}}{\w}
		&= -\left( \fracat{\partial \gamma(\theta, \w)}
                 	  {\partial \theta^T}{\thetahat(\w), \w} \right)^{-1}
			\fracat{\partial \gamma(\theta, \w)}{\partial \w^T}{\thetahat(\w), \w},
\end{align}
where we can take the inverse by our full-rank assumption.

We apply the general setting above to our special case with $\gamma(\theta, \w) =
\sumn w_n G(\theta, \d_{n})$ to find
\begin{align} \eqlabel{dtheta_dw}
	\fracat{\dee \thetahat(\w)}{\dee \w^T}{\w}
		&= -\left( \sumn w_n
        			\fracat{\partial G(\theta, \d_{n})}
               		{\partial \theta^T}{\thetahat(\w)} \right)^{-1}
			\left(
   				 G(\thetahat(\w), \d_{1}), \ldots, G(\thetahat(\w), \d_{N})
			\right),
\end{align}
which can again be computed with automatic differentiation software.

%% file: function_examples.tex
We end this section with some concrete examples of quantities of interest.
Recall from the start of \secref{MIP} that we are often interested in whether we
can change the sign or significance of an estimator, or generate a significant
result of the opposite sign. Recall that $\thetafun(\cdot)$ with only one
argument is a function of $\theta$, and $\thetafun(\cdot, \cdot)$ with two
arguments is a function of both $\theta$ and the weights $\w$.

To form our motivating examples, suppose for the remainder of this section we
are interested in the $\p$-th component of $\hat\theta$, where $\thetahat_{\p}$
is positive and statistically significant.  That is, let $\hat\sigma_{\p}$ be an
estimator of the variance of the limiting distribution of
$\sqrt{N}\thetafunhat$, and let $\thetahat_p - \frac{1.96}{\sqrt{N}}
\hat\sigma_{\p}$ be the lower end of our confidence interval. So we assume
$\thetahat_p > 0$ and $\thetahat_p - \frac{1.96}{\sqrt{N}} \hat\sigma_{\p} > 0$.
Moreover, we will write $\hat\sigma_{\p}(\theta, \w)$ to emphasize that standard
errors are typically given as functions of $\theta$ and the weights $\w$.  For
example, standard errors based on the observed Fisher information matrix $\meann
\w_n \fracat{\partial G(\theta, \d_n)}{\partial \theta}{\thetahat(\w)}$ will, in
general, depend on the weights both explicitly and through $\thetahat(\w)$.

To make $\thetahat_{\p}$ change sign, we can take
\begin{align} \eqlabel{function_change_sign}
\thetafun(\theta) =&
- \theta_{\p}.
& \textrm{(Change sign)}
\end{align}
We use $-\theta_{\p}$ instead of $\theta_{\p}$ since we have defined $\thetafun$
as a function that we are trying to increase (cf.\ \eqref{mis_weight} and the
discussion after). Increasing $\thetafun(\thetahat)$, for $\thetafun$ in
\eqref{function_change_sign}, by an amount $\Delta = \thetahat_{\p}$ is
equivalent to $\thetahat_{\p}$ changing sign from positive to negative.

To make $\thetahat_{\p}$ statistically non-significant, we wish
to take the lower bound of the confidence interval to $0$. To that end, we can take
\begin{align} \eqlabel{function_change_significance}
\thetafun(\theta, \w) =&
- \left(\theta_{\p} - \frac{1.96}{\sqrt{N}} \hat\sigma_{\p}(\theta, \w) \right).
& \textrm{(Change significance)}
\end{align}
As in the previous case, we choose \eqref{function_change_significance} with a
leading negative sign because we are trying to increase $\thetafun$ (cf.\
\eqref{mis_weight}). Increasing $\thetafun(\thetahat, \w)$, for $\thetafun$ in
\eqref{function_change_significance}, by an amount $\Delta = \thetahat_{\p} -
\frac{1.96}{\sqrt{N}} \hat\sigma_{\p}$ is equivalent to $\thetahat_{\p}$
becoming statistically insignificant.

Similarly,
to change to a significant result of the opposite sign, we can take
\begin{align*}
\thetafun(\theta, \w) =&
- \left(\theta_{\p} + \frac{1.96}{\sqrt{N}} \hat\sigma_{\p}(\theta, \w) \right)
& \textrm{(Significant sign reversal)}
\end{align*}
and $\Delta = \thetahat_{\p} + \frac{1.96}{\sqrt{N}} \hat\sigma_{\p}$, for if
the upper end of the confidence interval is negative, then the estimator must be
negative and statistically significant.

In each case above, the quantity $\Delta$ represents how far we must move
$\thetafun$ in order to reverse our conclusions.  In this sense, $\Delta$ is a
measure of the amount of ``signal'' in the original dataset.  As we will discuss
in \secref{why} below, the signal $\Delta$ is one of the three
key quantities that determine AMIP robustness.

%% file: introductory_regression_example.tex

Before continuing, we illustrate our method with an example. Economists often
analyze causal relationships using linear regressions estimated via ordinary
least squares (OLS), but a researcher rarely believes the conditional mean
dependence is truly linear. Rather, researchers use linear regression since it
allows transparent and straightforward estimation of an average treatment effect
or local average treatment effect. Researchers often invoke the law of large
numbers to  justify the focus on the sample mean, and invoke the central limit
theorem to justify the use of Gaussian confidence intervals when the sample is
large. We now discuss an example from recent economics literature showing how,
in practice, the omission of a very small number of data points can have outsize
influence on regression parameters in the finite sample even when the full
sample is large. We will study AMIP sensitivity for OLS further using simulation
and theory in \secref{influence_function_ols} below.

Consider as an example the set of seven randomized controlled trials of
expanding access to microcredit discussed by \citet{meager2019understanding}.
For illustrative purposes we single out the study with the largest sample size:
\citet{angelucci2015microcredit}. This study has approximately 16,500
households. A full treatment of all seven studies is in
\secref{example_microcredit_linear, example_microcredit_hierarchical} along with
tables and figures of the results discussed below.

We consider the headline results on household business profit regressed on an
intercept and a binary variable indicating whether a household was allocated to
the treatment group or to the control group. Let $Y_{ik}$ denote the profit
measured for household $i$ in site $k$, and let $T_{ik}$ denote their treatment
status. We estimate the following model via OLS with the regression formula
$Y_{ik} \sim \beta_0 + \beta_1 T_{ik}$. In the notation of
\secref{taylor_series}, we have $\theta = (\beta_0, \beta_1)^T$, $d_{ik} =
(Y_{ik}, T_{ik})$ with $n = (i, k)$, and $G(\theta, \d_{ik}) = (Y_{ik} -
(\beta_0 + \beta_1 T_{ik})) (1, T_{ik})^T$.

We confirm the main findings of the study in estimating a non-significant
average treatment effect (ATE) of -4.55 USD PPP per 2 weeks, with a standard
error of 5.88. We are interested in whether we can change the sign of $\beta_1$
from negative to positive, so we take $\thetafun(\theta) = \beta_1$. We compute
$\infl_n$ for each data point in the sample, which takes only a fraction of a
second in \textsf{R} using our Zaminfluence package.

Examining $\inflvec$, we find that one household has $\infl_n = 4.95$; removing
that single household should flip the sign if the approximation is accurate. We
manually remove the data point and re-run the regression, and indeed find that
the ATE is now 0.4 with a standard error of 3.19. Moreover, by removing 15
households we can generate an ATE of 7.03 with a standard error of 2.55: a
significant result of the opposite sign.

How is it possible for the absence of a single household to flip the sign of an
estimate that was ostensibly based on all the information from a sample of
16,500?  It may be tempting to suspect the use of sample means, which are known
to be non-robust to gross errors, or to speculate that such excess sensitivity
is simply symptomatic of ordinary sampling noise which is captured adequately by
standard errors.  In \secref{why} to follow, we show that such intuition is not
correct.  On the contrary, AMIP robustness is in fact fundamentally different
than both standard errors and classical robustness to gross errors.

%

%% file: theory_intro.tex
We now establish the determinants and accuracy of AMIP robustness. We begin by
deriving the key quantities of AMIP robustness in the simple case of correctly
specified univariate OLS regression (\secref{influence_function_ols}). For this
simple case, we show with theory and simulations that AMIP robustness is not
necessarily driven by misspecification, that AMIP non-robustness does not vanish
asymptotically, and that AMIP robustness is distinct from standard errors. Next,
we formally extend these conclusions to general Z-estimators in
\secref{influence_function}.  Finally, in \secref{accuracy}, we establish
conditions under which the approximation is provably uniformly accurate for
small $\alpha$, both in finite sample and asymptotically.

We will see that a central equation in our understanding of AMIP robustness is
its decomposition into three key quantities: the signal, noise, and shape.
First, the \emph{signal} $\Delta$ is the size of change in our quantity of
interest that would reverse our substantive conclusion (see
\secref{function_examples} above). Large values of the signal $\Delta$ indicate
that large changes are needed to make a different decision.  Second, the
\emph{noise} $\inflscale$ is defined by
\begin{align}\eqlabel{inflscale_def}
\inflscale^2 := \meann (N \infl_n)^2
\end{align}
We call $\inflscale$ the noise because $\inflscale^2$ is typically a consistent
estimator of the variance of the limiting distribution of $\sqrt{N}
\thetafun(\thetahat)$, a fact that will follow below from the relationship
between AMIP robustness, robust standard error estimators, and the influence
function (see \secpointref{amip_decomposition}{noise} or, more
generally, \secpointref{influence_function_for_real}{scale_via_influence}).
Third, the \emph{shape} $\shape$ is defined as
\begin{align}\eqlabel{shape_def}
\shape := -\frac{1}{N}
\sum_{n=1}^{\lfloor \alpha N \rfloor} \frac{N \infl_{(n)}}{\inflscale}
\ind{\infl_{(n)} < 0},
\end{align}
where $\infl_{(n)}$ refers to the $n$-th order statistic of the influence
scores, and $\ind{\cdot}$ denotes the indicator function taking value $1$ when
its argument is true and $0$ otherwise. The shape $\shape$ depends in a
complicated way on the shape of the tail of the distribution of the influence
scores, but we show that $0 \le \shape \le \sqrt{\alpha(1 - \alpha)}$ with
probability one, and that $\shape$ converges in probability to a nonzero
constant under standard assumptions (see
\secpointref{amip_decomposition}{shape}).  Given these three quantities, we will
show in \secpointref{amip_decomposition}{amip_decomposition} that
\begin{align} \eqlabel{robustness_three_parts}
\textrm{An analysis is AMIP non-robust }\quad\Leftrightarrow\quad
\frac{\Delta}{\inflscale} \le \shape.
\end{align}
We refer to the quantity $\Delta / \inflscale$ as the \emph{signal-to-noise ratio}.
For a given $\alpha$, \eqref{robustness_three_parts} suggests that it is the
signal-to-noise ratio that primarily determines AMIP robustness. Additionally, this
decomposition allows us to succinctly compare AMIP robustness to standard errors
and gross-error robustness, as well as to analyze the large-$N$ behavior of
AMIP robustness.

This section will use the following notation.  Let the symbol $\plim$ denote
convergence in probability, and $\dlim$ denote convergence in
distribution, both as $N \rightarrow \infty$.  Let $\vnorm{\cdot}_{op}$ denote
the operator norm of a matrix.

%% file: influence_ols_example.tex
We begin by focusing on the simple case of correctly-specified univariate linear
regression, both to provide intuition and motivate the more general results
that follow.

\subsubsection{Problem setup for Ordinary Least Squares example}

\point{Model}
\sloppy Let $X=(x_1, \ldots, x_N)^T$ denote a vector of $N$ continuous mean-zero
regressors, drawn IID from a distribution with finite variance $\sigma_x^2$. Let
$\varepsilon=(\varepsilon_1, \ldots, \varepsilon_N)$ be a vector of IID
draws from a $\mathcal{N}(0, \sigma_\varepsilon^2)$ distribution, where
we will assume $\sigma_\varepsilon$ is known.  For some unknown $\theta_0 \in
\mathbb{R}$, let $y_n = \theta_0 x_n + \varepsilon_n$, so that the vector
$Y=(y_1, \ldots, y_N)$ given $X$ is drawn from a correctly specified regression
model with true coefficient $\theta_0$.

\point{Weighted estimating equation}
The OLS estimator $\thetahat$ is traditionally found by maximizing the (log)
likelihood: $\log p(y_n \vert \theta, x_n) = -\frac{1}{2
\sigma_\varepsilon^{2}}(y_n - \theta x_n)^2 + C$, where $C$ does not depend on
$\theta$. In particular, setting the derivative of the log likelihood to zero
yields the estimating equation $G(\theta, \d_n) =
-\frac{1}{\sigma_\varepsilon^{2}} (y_n - \theta x_n) x_n = 0$. That is,
$\thetahat$ is a Z-estimator with this choice of $G$ (see
\eqref{estimating_equation_no_weights}). Typical Z-estimators do not have
closed-form solutions. But in this case, the solution to the estimating equation
returns the usual OLS estimate. A similar derivation returns the solution to the
weighted estimating equation given in \eqref{estimating_equation_with_weights}:
$\thetahat(\w) = \left(\meann \w_n x_n^2 \right)^{-1} \meann \w_n y_n x_n$.

\point{Quantity of interest}
Suppose we are interested in the sign of
$\theta_0$. Without loss of generality, we assume $\thetahat < 0$. Then our
quantity of interest is $\thetafun(\theta) = \theta$.

\point{Signal and noise}
\pointlabel{ols_signal_and_noise}
For our quantity of interest, the signal is $\Delta = \abs{\thetahat}$ since, if
we can increase $\thetahat$ by an amount $\abs{\thetahat}$, its sign will
change. To compute the noise, we compute the influence scores. Directly
differentiating the explicit formula for $\thetahat$ gives, as it must, the same
value for $\infl_n$ as the implicit function theorem result of
\eqref{dtheta_dw_general}. Letting $\hat\varepsilon_n := y_n - \thetahat x_n$
and $S_X := \meann x_n^2$, we see, either by direct differentiation or by
\eqref{dtheta_dw_general}, that $\infl_n = N^{-1} S_X^{-1} x_n
\hat\varepsilon_n$. For intuition about the noise $\inflscale$, we observe its
asymptotic behavior. Standard results for OLS give:
\begin{align} \eqlabel{ols_limit_of_noise}
\inflscale^2 = \meann (N \infl_n)^2 = S_X^{-2} \meann x_n^2 \hat\varepsilon_n^2
\plim \frac{\sigma_\varepsilon^2}{\sigma_x^2}.
\end{align}
Note that the noise includes a contribution from both the residual and regressor
variance---we describe $\inflscale$ as the ``noise'' because it estimates the
variability of $\sqrt{N} \thetahat$, not of the residuals (see
\secpointref{ols_what_determines}{ols_se_not_amip} below). Finally, we emphasize
that, although we will be using asymptotics to provide intuition, by ``noise''
we will always mean the finite-sample quantity $\inflscale$, not its asymptotic
limit.

\subsubsection{What determines AMIP robustness for Ordinary Least Squares?}
\seclabel{ols_what_determines}
Now that we have translated OLS into our framework, we can
analyze the AMIP for OLS. To that end, we use both theory and a simulation study.
We outline the simulation study before describing our main conclusions.
For $N=\SimNumObs$ data points, and for a range of $\sigma_x$ and
$\sigma_\varepsilon$, we drew normal regressors $x_n \sim \mathcal{N}(0,
\sigma_x^2)$ and residuals $\varepsilon_n \sim \mathcal{N}(0,
\sigma_\varepsilon^2)$. For $\theta_0 =
\SimTrueTheta$, we set $y_n = \theta_0 x_n + \varepsilon_n$.
We computed the OLS estimator $\hat\theta = \sumn y_n x_n /
\sumn x_n^2$.

\SimCombNormalGraph{}

\point{Signal-to-noise ratio drives AMIP robustness}
From our discussion at the start of \secref{why}, we expect that the
signal-to-noise ratio drives whether an analysis is AMIP-robust or not. In our
simulation, $N$ is large and we keep $\theta_0$ fixed, so we expect that the
signal does not change substantially over the simulation. Therefore,
signal-to-noise is controlled by the noise. Following the asymptotic argument
above, we approximate the noise as $\sigma_\varepsilon / \sigma_x$. In the left
panel of \figref{sim-comb-normal}, we vary $\sigma_\varepsilon$ and $\sigma_x$
and plot the resulting Approximate Perturbation Inducing Proportion $\alpha^*$
to change the sign of $\hat\theta$. As expected, we see that the simulations
with the largest approximate noise $\sigma_\varepsilon / \sigma_x$ are the least
robust, in the sense that one can reverse the sign of $\hat\theta$ by dropping a
very small proportion of points.

\point{Influential data points have both a large residual and large regressor}
Let $(\hat\varepsilon x)_{(n)}$ denote the products $\hat\varepsilon_n x_n$,
sorted from most negative to most positive, so that the sorted influence scores
are $\infl_{(n)} = N^{-1} S_X^{-1} (\hat\varepsilon x)_{(n)}$.  From this
formula, we observe that influential datapoints have both a large residual and a
large regressor (relative to the regressor variance).\footnote{ Indeed, if we
had taken $\thetafun(\theta) = \thetahat x_n = \hat{y}_n$, then the $n$-th
influence score would have been $S_X^{-1} x_n^2 \hat\varepsilon_n$, which is
precisely the leverage score times the residual. This expression formalizes the
conceptual link made by \citet{chatterjee1986influential} between influence,
leverage, and large values of $\hat\varepsilon_n$. } A typical influence score
goes to zero at rate $N^{-1}$, though extreme values such as $\max_{n}
\abs{\infl_n}$ may obey a different rate.  However, since $\meann x_n^2$ and
$\meann \varepsilon_n^2$ are finite with high probability, even $\max_{n}
\abs{\infl_n}$ does not diverge in this case.\footnote{The finiteness follows
from the inequality $\frac{1}{N} \max_{n} x_n^2 \le \meann x_n^2 \plim
\sigma_x^2$, with an analogous inequality for $\varepsilon_n$. However, since we
know $\varepsilon_n$ is Gaussian, we actually have a stronger result in this
case: $\max_{n \in \{1,\ldots,N\}} \abs{\varepsilon_n}$ grows at rate
$\sqrt{\log (2N)}$ \citep[Theorem 1.14]{rigollet:2015:highdimstats}.}

\point{AMIP sensitivity does not vanish as $N \rightarrow \infty$}
Standard results for OLS give that $S_X \plim \sigma_x^2$ and $\hat\varepsilon_n -
\varepsilon_n \plim 0$. So $N \infl_n - \sigma_x^{-2} x_n \varepsilon_n \plim
0$. Consequently, the empirical distribution of $N \infl_n$ converges to a
non-degenerate distribution with finite variance.  Let $q_\alpha$ denote the
$\alpha$-th quantile of the distribution of the random variable $\sigma_x^{-2}
x_1 \varepsilon_1$.  Since $x_n$ and $\varepsilon_n$ are independent, and about
half of the $\varepsilon_n$ will be negative, we expect about half of the
influence scores to be negative. So for $\alpha \ll 1/2$, with high probability
at least $\alpha N$ influence scores are negative. Then, by \eqref{w_approx_opt}
and Slutsky's theorem, we have
\begin{align*}
\thetafunlin(\w^*) -
\thetafunhat
= -\sum_{n=1}^{\alpha N} \infl_{(n)}
= - \frac{1}{ N}
\sum_{n=1}^{\alpha N} S_X^{-1} (\hat\varepsilon x)_{(n)}
\plim
\expect{-\frac{x_1 \varepsilon_1}{\sigma_x^2}
               \ind{\frac{x_1 \varepsilon_1}{\sigma_x^2} \le q_\alpha}}.
\end{align*}
The right hand side of the preceding display is strictly positive for finite
$\alpha$.  So, for fixed $\alpha$, we expect that AMIP sensitivity does not
vanish as $N \rightarrow \infty$.\footnote{As desired, though, the expectation
does go to zero as $\alpha \rightarrow 0$ since $\expect{\abs{x_1
\varepsilon_1}} < \infty$.}

\point{AMIP non-robustness is not due only to misspecification}
Our simulations are well specified. Yet we see from \figref{sim-comb-normal}
that different cases can still be robust or non-robust under various robustness
cut-offs---according to their differing signal-to-noise ratios.

Asymptotically as $N \rightarrow \infty$, even in a well-specified model, we in
fact expect AMIP non-robustness at any $\alpha$ for a sufficiently small
$\abs{\theta_0}$. The limiting value of the AMIP sensitivity does not depend on
$\theta_0$.  Thus, as $N \rightarrow \infty$, our quantity of interest (for
changing the sign of the estimator) will be AMIP non-robust with high
probability if and only if $\abs{\theta_0} < \expect{-\frac{x_1
\varepsilon_1}{\sigma_x^2} \ind{\frac{x_1 \varepsilon_1}{\sigma_x^2} \le
q_\alpha}}$.  If we are interested in the sign of $\theta_0$, and
$\abs{\theta_0}$ is small relative to the tail means of $\sigma_X^{-2} x_1
\varepsilon_1$, then the problem will be AMIP non-robust with probability
approaching one, no matter how large $N$ is---despite the fact that the model is
correctly specified and there are no abnormalities in the data.

\point{Though both are scaled by the noise, standard errors are different
from---and typically smaller than---AMIP sensitivity}
\pointlabel{ols_se_not_amip}
In what may seem at first like a remarkable coincidence, the variance of the
limiting distribution of $N \infl_n$ (which determines AMIP sensitivity---see
\eqref{taylor_approx}) is the same as the variance of the limiting distribution
of our quantity of interest $\sqrt{N}(\thetahat - \theta_0)$ (which determines
classical standard errors). The two distributions are not the same---the
limiting distribution of $N \infl_n$ is not, in general, normal---but they
have the same scale.  In particular, compare the noise limit in
\eqref{ols_limit_of_noise} with the following limit, which follows by standard
results for OLS.
\begin{align*}
\quad
\sqrt{N}(\thetahat - \theta_0) \dlim
\mathcal{N}\left(0, \frac{\sigma_\varepsilon^2}{\sigma_x^2} \right).
\end{align*}
As we discuss below in \secpointref{amip_decomposition}{noise}
and \secpointref{influence_function}{scale_via_influence}, this equality is no
coincidence, but a general (and well-known) relationship between influence
scores and the limiting distributions of quantities of interest.

For large $N$, use of standard errors will admit the hypothesis that $\theta_0$
might be $0$ whenever $\abs{\theta_0} < \frac{1.96}{\sqrt{N}}
\frac{\sigma_\varepsilon}{\sigma_x}$. Thus, for every $\theta_0 \ne 0$, using
standard errors always rejects $\theta_0 = 0$ for sufficiently large $N$.
By contrast, as we saw above, using the AMIP will admit a change large enough to
move $\thetahat$ to $0$ whenever
\begin{align*}
\abs{\theta_0} \le
\left(
\expect{-\frac{x_1}{\sigma_x} \frac{\varepsilon_1}{\sigma_\varepsilon}
   \ind{\frac{x_1 }{\sigma_x} \frac{\varepsilon_1}{\sigma_\varepsilon}
    \le \frac{\sigma_x}{\sigma_\varepsilon} q_\alpha
   }}
   \right)
   \frac{\sigma_\varepsilon}{\sigma_x}
   \ne \frac{1.96}{\sqrt{N}}
   \frac{\sigma_\varepsilon}{\sigma_x}.
\end{align*}
Thus, we see that both the AMIP sensitivity and standard errors admit larger
possible values for $\thetahat$ when the limiting value $\abs{\theta_0} /
(\sigma_\varepsilon / \sigma_x)$ of the signal-to-noise ratio is large. But
AMIP sensitivity is determined by the tail mean of the standardized influence
scores, and standard errors are determined by a quantity that goes to zero as $N
\rightarrow \infty$. Thus AMIP sensitivity is distinct from, and typically
larger than, standard errors. The tail behavior of the unit-variance random
variable $\frac{x_1}{\sigma_x} \frac{\varepsilon_1}{\sigma_\varepsilon}$ is
exactly the shape we introduced at the start of \secref{why}. The shape captures
the scale-independent shape of the tails of the distribution of the influence
scores; see \secpointref{amip_decomposition}{shape} below for a detailed and
general analysis.

\point{Our approximation is accurate for small $\alpha$}
\pointlabel{ols_small_alpha}
The expression for
$\thetahat(\w)$ depends on two terms, $\left(\meann \w_n x_n^2 \right)^{-1}$ and
$\meann \w_n y_n x_n$, both of which are uniformly smooth functions of $\w / N$
with high probability for sufficiently small $\vnorm{\w - \onevec}_2 / N$.  As a
consequence of smoothness, we expect a linear approximation formed at $\w =
\onevec$ to be accurate when $\vnorm{\w - \onevec}_2 / N$ is small.  And when
$\w$ contains no more than $\lfloor \alpha N \rfloor$ zeros and the rest ones,
we have that $\vnorm{\w - \onevec}_2 / N \le \alpha$, so we expect a linear
approximation to be accurate when $\alpha$ is small. We make this intuition
precise and general in \secref{accuracy} below.

We check the accuracy of the approximation empirically in
\figref{sim-comb-normal}. For the right hand plot in \figref{sim-comb-normal},
we fixed $\sigma_\varepsilon = \SimAccSigeps$ and $\sigma_x = \SimAccSigx$. We
computed the Approximate Most Influential Set for a range of left-out
proportions $\alpha$ from $0$ to $\SimAccPercentMax \%$.  For each $\alpha$, we
computed the linear approximation, re-ran the regression to compute the actual
change, and computed the error of the linear approximation as the difference of
the two.  The right panel of \figref{sim-comb-normal} shows how the relative
error of the approximation vanishes for small $\alpha$, and that, qualitatively,
the approximation is very good for removal proportions less than $2.5\%$.

%% file: influence_function.tex
\def\tz{T_Z}
\def\tfun{T_\thetafun}
\def\fhat{\hat{F}_N}
\def\flim{F_{\infty}}
\def\falt{F_{alt}}
\def\fbase{F_{0}}
\def\ic{\mathrm{IF}}
\def\ichat#1{\widehat{IC}_{#1}}
\def\wsum{N_{\w}}

We next show that the conclusions of \secref{influence_function_ols} hold not
just for OLS but in considerable generality for Z-estimators applied to
IID data. In the present section, we will
establish more generally that AMIP sensitivity is not a product of
misspecification, does not vanish as $N$ goes to infinity, and is distinct from
standard errors. To that end, in \secref{amip_decomposition} we first formally
decompose the AMIP into the shape and noise terms defined at the beginning of
\secref{why}, and we establish that the shape is roughly constant across
distributions. Then, in \secref{amip_robustness_breakdown}, we use this
decomposition to revisit our OLS conclusions about AMIP sensitivity but now more
broadly. Finally, in \secref{influence_function_for_real}, we connect the AMIP
to the influence function, showing how AMIP robustness is different from gross
error robustness.

\subsubsection{The decomposition of the AMIP}
\seclabel{amip_decomposition}

\point{The AMIP is the noise times the shape}
\pointlabel{amip_decomposition}
Let $\infl_{(1)}, \ldots, \infl_{(N)}$ denote the order statistics of the
influence scores.  Recall that the Approximate Maximum Influence Perturbation is
given by the negative of the sum of the $\lfloor \alpha N \rfloor$ largest
influence scores.  So we can write
\begin{align}
\amip{\alpha} = \thetafunlin(\w^{*}) - \thetafunhat =
- \sum_{n=1}^{\lfloor \alpha N \rfloor} \infl_{(n)} \ind{\infl_{(n)} < 0} =
\inflscale \shape.
\end{align}
The first equality follows from the definition of the AMIP $\amip{\alpha}$
(\defref{approx_metrics}). The second equality follows from
\eqref{w_approx_opt}. The third equality follows from the definitions of noise
$\inflscale$ and shape $\shape$ at the start of \secref{why}.

\point{The noise is an estimator of the standard deviation of the limiting
distribution of the quantity of interest (Z-estimator version)}
\pointlabel{noise}
In the case of Z-estimators, we can show by direct computation that
$\inflscale^2$ is the estimator of the variance of the limiting distribution of
$\sqrt{N}\thetafun(\thetahat)$ given by the delta method and the ``sandwich'' or
``robust'' covariance estimator \citep{huber1967sandwich,
stefanski:2002:mestimation}.  To see this, observe first that $\meann
\fracat{\dee \thetahat(\w)}{\dee \w_n}{\onevec} \left( \fracat{\dee
\thetahat(\w)}{\dee \w_n}{\onevec}\right)^T$, as given by \eqref{dtheta_dw}, is
precisely the sandwich covariance estimator for the covariance of the limiting
distribution of $\sqrt{N} \thetahat$.  In turn, the sample variance of the
linear approximation given in \eqref{chain_rule_influence_score}, given by
$\inflscale^2$, is then the delta method variance estimator for
$\sqrt{N}\thetafunhat$.  Note that we came to the same conclusion in the special
case of OLS in \secpointref{ols_what_determines}{ols_se_not_amip} above.

It follows that we can use $\inflscale$ to form consistent credible intervals
for $\thetafun$, a fact that will be useful below when comparing AMIP robustness
to standard errors.  Specifically, if $\inflscale \plim \inflscalelim$ and
$\thetahat \plim \thetalim$, then
\begin{align}\eqlabel{z_normal_limit}
\sqrt{N}(\thetafun(\thetahat) -
\thetafun(\thetalim)) \dlim \mathcal{N}(0, \inflscalelim^2).
\end{align}
As we discuss in \secpointref{influence_function_for_real}{scale_via_influence}
below, this relationship between asymptotic variance and the influence scores
is in fact a consequence of a general relationship between influence functions
and distributional limits.

\point{The shape depends primarily on $\alpha$, not on the model specification}
\pointlabel{shape}
More precisely, we next show that the shape $\shape$ satisfies the following
properties. (1) With probability one, $0 \le \shape \le
\sqrt{\alpha(1-\alpha)}$. (2) Typically, $\shape$ converges in probability to a
nonzero constant as $N \rightarrow \infty$. (3) $\shape$ is largest when the
influence scores of the left-out points are all equal. Conversely, heavy tails
in the distribution of $\infl_n$ result in smaller values of $\shape$. (4)
Empirically, $\shape$ varies relatively little among common sampling
distributions.

To prove the lower bound in (1), we observe that the indicator $\ind{\infl_{(n)} <
0}$ accounts for the fact that the adversarial weight would leave out fewer
points rather than drop a point with positive $\infl_{(n)}$.  Because of this,
$\shape \ge 0$. We show the upper bound of (1) as part of the extremization
argument for (3) below.

To prove (2), notice that $\shape$ is a sum of $\lfloor \alpha N \rfloor$
positive terms, divided by $N$. In general, then, we expect $\shape$ to converge
to a nonzero constant for fixed $\alpha$ as long as the distribution of $N
\infl_n$ converges marginally in distribution to a non-degenerate random
variable.  And indeed, by \eqref{chain_rule_influence_score, dtheta_dw}, we
expect such convergence from Slutsky's theorem as long as $\thetahat$ and
$\meann \fracat{\partial G(\thetahat, \d_n)}{\partial\theta}{\thetahat}$
converge in probability to constants, since $N \infl_n$ is proportional to
$G(\thetahat, \d_n)$, which itself has a non-degenerate limiting distribution.

We next show (3), that $\shape$ takes its largest possible value when all the
influence scores $\infl_{(1)}, \ldots, \infl_{(\alpha N)}$ take the same
negative value. To that end, take $\alpha N$ to be an integer for simplicity. By
the definition of $\inflscale$ (\eqref{inflscale_def}), $\meann \left( \frac{N
\infl_{(n)}}{\inflscale} \right)^2 = 1$, and by properties of the influence
function detailed below, $\sumn \infl_n = 0$
(\secpointref{influence_function_for_real}{infl_sum_zero}). So $\shape$ is a
tail average of scalars with zero sample mean and unit sample variance.
Therefore, it is equivalent to consider scalars $z_1, \ldots, z_N$ with $\meann
z_n = 0$ and $\meann z_n^2 = 1$ and to ask how to maximize the average
$-\frac{1}{\alpha N} \sum_{n=1}^{N\alpha} z_{(n)}$.

To perform this maximization we divide datapoints into a set $D$ of dropped
indices, and set $K$ of kept indices. To be precise, $D := \{n: z_{(n)} \le
z_{(\alpha N)} \}$ and $K := \{1,\ldots,N\} \setminus D$. We write the sample
means and variances within the sets respectively as $\mu_D := \frac{1}{\alpha N}
\sum_{n \in D} z_n$ and $v_D := \frac{1}{\alpha N} \sum_{n \in D} (z_n -
\mu_D)^2$, with analogous expressions for $\mu_K$ and $v_K$. In this notation,
our goal is to extremize $\mu_D$, the mean in the dropped set.   The constraints
on the distribution can then be written as $\meann z_n = 0 \Rightarrow \alpha
\mu_D + (1- \alpha) \mu_K = 0$, and $\meann z_n^2 = 1 \Rightarrow \alpha(v_D +
\mu_D^2) + (1 - \alpha) (v_K + \mu_K^2)  = 1$. Given these constraints, we
extremize $\mu_D$ by setting $v_K = v_D = 0$, in which case we achieve $\mu_D =
-\sqrt{(1 - \alpha) / \alpha}$. Identifying $N \infl_n / \inflscale$ with $z_n$,
and $\shape$ with $\alpha \mu_D$, we see that the worst-case value of $\shape$
occurs when all the influence scores $\infl_{(1)}, \ldots, \infl_{(\alpha N)}$
take the same negative value. This observation completes our argument for (3).
It also follows from this argument that $\shape \le \sqrt{\alpha (1 - \alpha)}$
with probability one, a bound that is achieved in the worst-case. This
observation supplies the upper bound in (1).

To establish point (4), we fix a representative $\alpha$, simulate a large
number of IID draws $\tilde{z}_n$ from some common distributions, standardize to
get $z_n := \frac{\tilde{z}_{n} - \bar{\tilde{z}}}{\sqrt{\meann (\tilde{z}_n -
\bar{\tilde{z}})^2}}$, and compute the shape $\shape = -\frac{1}{N}
\sum_{n=1}^{\lfloor \alpha N \rfloor} z_{(n)}$. We find that, across common
distributions, $\shape$ varies relatively little. For example, for $\alpha =
0.01$, a Normal distribution gives $\shape = 0.0266$, a Cauchy distribution
gives $\shape = 0.0022$.  As expected based on the reasoning of the previous
paragraph, the heavy-tailed Cauchy distribution has a smaller shape than the
Normal distribution.  The worst-case distribution, for which all left-out $z_n$
are equal, gives $\shape = 0.0995 \approx \sqrt{\alpha(1-\alpha)}$ as expected.

\subsubsection{What determines AMIP robustness?}
\seclabel{amip_robustness_breakdown}
We now use the decomposition of the AMIP into noise and shape, and the relative
stability of the shape, to derive a number of general properties of AMIP
robustness.

\point{Signal-to-noise ratio drives AMIP robustness}
\pointlabel{snr_drives_amip}
We argued above that we do not expect $\shape$ to vary radically from
one problem to another. By contrast, the noise $\inflscale$ can, in principle,
be any positive number.  We conclude then, that the signal-to-noise ratio,
rather than the shape, principally determines AMIP robustness.

This relationship also suggests what might be done if the analysis is deemed
AMIP non-robust. Since, as we showed in \secpointref{amip_decomposition}{noise},
$\inflscale$ is thus the same quantity that enters standard error computations,
analysts are typically attentive to choosing estimators with $\inflscale$ as
small as possible while still guaranteeing desirable properties like
consistency. Meanwhile, the signal $\Delta$ is determined by the question being
asked and the true state of nature as estimated by $\thetahat$.  In light of
these observations, consider a case where $\Delta / \inflscale$ is too small to
ensure AMIP robustness. Then it seems necessary for the investigator to ask a
different question, or investigate different data, to find an AMIP robust
analysis.

\point{AMIP sensitivity does not vanish as $N \rightarrow \infty$}
\pointlabel{amip_does_not_vanish}
Both $\inflscale$ and $\shape$ converge to nonzero constants. So $\inflscale
\shape$, the estimated amount by which you can change an estimator, does not go
to zero, either. If the signal $\Delta$ is less than the probability limit of
$\inflscale \shape$, then the problem will be AMIP non-robust no matter how
large $N$ grows. As we discuss below, this behavior contrasts sharply with the
behavior of standard errors.

\point{AMIP non-robustness is not due only to misspecification}
Consider a correctly-specified problem with no aberrant data points. As we
discussed above in  \secpointref{amip_decomposition}{noise},
the noise will still have some non-zero probability limit. We showed in
\secpointref{amip_decomposition}{shape} that the shape will have a non-zero
probability limit. And the quantity of interest $\thetafun(\thetahat)$ can
generally be expected to have a non-zero probability limit. So by the
decomposition of \eqref{robustness_three_parts}, if the user is interested in a
question whose signal is small enough, their problem will be AMIP non-robust,
despite correct specification.

\point{Though both are scaled by noise, standard errors are different from---and
typically smaller than---AMIP sensitivity} \pointlabel{amip_is_not_se}
Recall that classical standard errors based on limiting normal approximations
also depend on $\inflscale$, in that we typically report a confidence interval
for $\thetafun$ of the form
$\thetafun \in \left(\thetafun(\theta, \onevec) \pm q_{\mathcal{N}}
\frac{\inflscale}{\sqrt{N}}  \right)$,
where $q_{\mathcal{N}}$ is some quantile of the normal distribution, e.g. the
0.975-th quantile $q_{\mathcal{N}} \approx 1.96$.  In this sense, using standard
errors errors allow that $\thetafun$ may be as large as $\thetafun + \Delta$
whenever $\Delta / \inflscale \le \frac{1.96}{\sqrt{N}}$. By contrast, AMIP
robustness allows that $\thetafun$ may be as large as $\thetafun + \Delta$ when
$\Delta / \inflscale \le \shape$. Since $\shape \ne \frac{1.96}{\sqrt{N}}$ in
general, these two approaches will yield different conclusions.  Indeed,
typically $\shape$ converges to a non-zero constant as $N \rightarrow 0$, while
$\frac{1.96}{\sqrt{N}}$ converges to zero.

\point{Statistical non-significance is always AMIP-non-robust as $N
\rightarrow \infty$}
This observation follows as a corollary of the discussion above. In particular,
we might conclude statistical non-significance if $\abs{\thetafun(\thetahat,
\onevec)} \le \frac{1.96 \inflscale}{\sqrt{N}}$. To produce a statistically
significant result, and so undermine the conclusion, it suffices to move
$\thetafun(\thetahat, \onevec)$ by more than $\frac{1.96 \inflscale}{\sqrt{N}}$.
Take any $\alpha$. As we have seen above, we can produce a change of $\inflscale
\shape$, which is greater than $\frac{1.96 \inflscale}{\sqrt{N}}$ whenever
$\shape > 1.96 / \sqrt{N}$.  Thus, for any fixed $\alpha$, there always exists a
sufficiently large $N$ such that statistical non-significance can be undermined
by dropping at most $\alpha$ proportion of the data. By contrast, statistical
significance can be robust if $\thetafun(\thetahat, \onevec)$ converges to a
value sufficiently far from $0$.

\subsubsection{The influence function}
\seclabel{influence_function_for_real}

We next review the influence function, its known properties, and its particular
form for Z-estimators \citep[e.g.,][chapter 2.3]{hampel1986robustbook}. We first
show the relationship between the influence scores and the empirical influence
function. We use these connections to further justify the relationship between
the noise and the limiting distribution of $\sqrt{N}\thetafunhat$. Finally, we
use these classical properties of the influence function to contrast AMIP
robustness with gross error robustness and establish that outliers primarily
affect AMIP robustness via the noise, rather than via the shape.

\point{Writing a statistic as a functional of the empirical distribution}
Before defining the influence function, we set up some useful notation. Suppose
we observe IID data, $\d_1, \ldots, \d_N$. Each point is drawn from a data
distribution $\flim(\cdot) = p(\d_1 \le \cdot)$, where the inequality may be
multi-dimensional. For a generic distribution $F$, let $T$ represent a
functional of the distribution: $T(F)$.  One example is the sample mean; for a
generic distribution $F$, let $T_{mean}(F) = \int \tilde{\d} \dee
F(\tilde{\d})$.  Then $T_{mean}(\flim) = \expect{\d_1}$ is the population mean.
If we let $\fhat$ denote the empirical distribution function $\fhat(\cdot) =
\meann \ind{\cdot \le \d_n}$, then $T_{mean}(\fhat) = \meann \d_n$ is the sample
mean.

Now consider Z-estimators. Define $T_Z(F)$ to be a quantity satisfying
\begin{align} \eqlabel{estimating_equation_F}
	\int G(T_Z(F), \tilde{\d}) \dee F(\tilde{\d}) &= 0.
\end{align}
See, e.g., \citet[Section 4.2c, Def. 5]{hampel1986robustbook}. If we plug in
$\fhat$ for $F$ in \eqref{estimating_equation_F} (and multiply both sides by $N$),
we recover the Z-estimator estimating equation from
\eqref{estimating_equation_no_weights}, with solution $\thetahat = T_Z(\fhat)$.
Similarly, let $\hat{F}_w$ to be the distribution function putting weight
$N^{-1} w_n$ at data point $\d_{n}$. Plugging in $\hat{F}_w$ for $F$ in
\eqref{estimating_equation_F} yields the estimating equation in
\eqref{estimating_equation_with_weights}, for weighted Z-estimators, with
solution $\thetahat(\w) = T_Z(\hat{F}_w)$.
Finally, we can define a new functional $T_\thetafun(F)$ by applying the smooth
function $\thetafun$, which picks out our quantity of interest, to $T_Z(F)$:
$T_\thetafun(F) = \thetafun(T_Z(F), \onevec)$.\footnote{ As in ordinary calculus
in Euclidean space, we can also allow for explicit $F$ dependence in $\thetafun$
by writing $\thetafun(\theta, F)$. Allowing this level of generality, though, is
notationally burdensome and not typical in the analysis of the influence
functions for Z-estimators. So we omit this dependence for simplicity.}

\point{The influence function}
The influence function $\ic(\d; T, F)$ measures the effect on a statistic
$T$ of adding an infinitesimal amount of mass at point $\d$ to some base or
reference data distribution $F$
\citep{reeds1976thesis,hampel1986robustbook}. Let $\delta_\d$ be the probability
measure with an atom of size $1$ at $\d$. Then
\begin{align}\eqlabel{influence_function_def}
\ic(\d; T, F) := \lim_{\epsilon \searrow 0} \frac{ T(\epsilon \delta_\d +
(1-\epsilon) F) - T(F)}{\epsilon}.
\end{align}
The influence function is defined in terms of an ordinary univariate derivative,
and can be computed (as a function of $\d$ and $F$) using standard univariate
calculus.  In particular, our quantity of interest has the following influence
function:
\begin{align}
\eqlabel{influence_function_Z_estimator}
	\ic(\d; T_\thetafun, F)
	&=
		-\fracat{
				\partial \thetafun(\theta, \onevec)
			}{
				\partial \theta^T
			}{
				\thetahat(F)
			}
		\left(\int
       			\fracat{
					\partial G(\theta, \tilde{\d})
				}{
					\partial \theta^T
				}{
					\thetahat(F)
				}
			\dee F(\tilde{\d})
    		\right)^{-1}
    		G(\thetahat(F), \d).
\end{align}
By comparing \eqref{influence_function_Z_estimator} with
the definition of $\infl_n$ in \eqref{chain_rule_influence_score, dtheta_dw}, we can see
that, formally,\footnote{The factor of $N$ arises to re-write the expectation
as a sum over unit-valued weights.}
\begin{align}\eqlabel{infl_is_infl}
N \infl_n = \ic(\d_n; T_\thetafun, \fhat).
\end{align}
\Eqref{infl_is_infl} is not a coincidence.  To see this, note that the set of
distributions that can be expressed as weighted empirical distributions
($\hat{F}_w$ above) is precisely the subspace of possible distribution functions
concentrated on the observed data.  So the derivative $N \infl_n = N \partial
\thetafun(\thetahat(\w), \onevec) / \partial \w_n$ (\eqref{taylor_approx}) is
simply a path derivative representation of the functional derivative $\ic(\d_n;
T_\thetafun, \fhat)$.

We refer to the influence function applied with $F = \fhat$ as the
\emph{empirical influence function} \citep{hampel1986robustbook}. We conclude
that the $\infl_n$ that we use to form our approximation are the values of the
empirical influence function at the datapoints $\d_1, \ldots, \d_N$. For this
reason, we refer to the $\infl_n$ as influence scores.

\point{The sum of the influence scores is zero}
\pointlabel{infl_sum_zero}
We can now use standard properties of the influence function to reason about
$\inflvec$. For instance, the fact that $\sumn \infl_n = 0$ follows from
\eqref{influence_function_Z_estimator} and the fact that $\thetahat$ solves
\eqref{estimating_equation_no_weights}.

\point{The noise is an estimator of the standard deviation of the limiting
distribution of the quantity of interest (influence function version)}
\pointlabel{scale_via_influence}
Observe that, by our influence function development above, we can write
the squared noise as follows.
\begin{align} \eqlabel{scale_is_influence_norm}
\inflscale^2 := N \vnorm{\inflvec}_2^2 =
\meann (N \infl_n)^2 = \meann \ic(\d_n; T_\thetafun, \fhat)^2,
\end{align}

Recall that we saw above that $\inflscale^2$ consistently estimates the variance
of the limiting distribution of $\sqrt{N}\thetafunhat$, first in the special
case of OLS (\secpointref{ols_what_determines}{ols_se_not_amip}) and then for
Z-estimators in general (\secpointref{amip_decomposition}{noise}).  We can now
see that those results are themselves special cases of the following well-known
relationship between the influence function and the limiting variance of its
corresponding functional:
\begin{align}\eqlabel{infl_normal_limit}
\sqrt{N}\left(T(\fhat) - T(\flim) \right) \dlim
\mathcal{N}\left(0, \expect{\ic(\d_1; T, \flim)^2}\right),
\end{align}
where the expectation in the preceding display is taken with respect to $\d_1
\sim \flim$ (see, e.g., \citet[Eq.
2.1.8]{hampel1986robustbook}).\footnote{Though \eqref{infl_normal_limit} can
provide useful intuition, as it does in our case, it is often easier in any
particular problem to prove asymptotic results directly rather than through the
functional analysis perspective of this section, since stating precise and
general conditions under which \eqref{infl_normal_limit} holds can be
challenging. See, for example, the discussion in \citet[Chapter
6]{serfling2009approximation} or \citet[Chapter 20]{vaart2000asymptotic}.}
Specifically, if we can show that $\inflscalelim$, the probability limit of
$\inflscale$, is equal to $\expect{\ic(\d_1; T, \flim)^2}$, then
\eqref{infl_normal_limit} would imply $\sqrt{N}(T_\phi(\fhat) - T_\phi(\flim))
\dlim \mathcal{N}(0, \inflscalelim^2)$, just as we showed in
\eqref{z_normal_limit} using the sandwich covariance estimator. In our case,
under standard assumptions, one can show directly from
\eqref{chain_rule_influence_score, dtheta_dw} that $\ic(\d_n; T_\thetafun,
\fhat) \plim \ic(\d_n; T_\thetafun, \flim)$, almost surely in $\d_n$.  A law of
large numbers can then be applied to \eqref{scale_is_influence_norm} giving the
desired result.

\point{AMIP robustness is different from gross error robustness}
\pointlabel{gross_errors}
Roughly speaking, an estimator is considered non-robust to gross errors if its
influence function is unbounded \citep{huber1981robust}.  For instance, the
influence function arising from the OLS Z-estimator
(\secref{influence_function_ols}) is classically known to be non-robust to gross
errors. When an influence function is unbounded, one can produce arbitrarily
large changes in the quantity of interest by making arbitrarily large changes to
a single datapoint.  Gross-error robustness is motivated by the possibility that
some small number of datapoints come from a distribution arbitrarily different
from the model's posited distribution. By contrast, to assess AMIP robustness,
we do not make arbitrarily large changes to datapoints. We simply remove
datapoints. And the analysis is AMIP-non-robust if a change of a particular size
($\Delta$) can be induced, rather than an arbitrarily large change.
Consequently, problems with unbounded influence functions (such as OLS in
\secref{influence_function_ols}) can be AMIP-robust if $\Delta / \inflscale$ is
sufficiently large. And perfectly specified problems with no outliers can be
AMIP non-robust if $\Delta / \inflscale$ is sufficiently small.

\point{Outliers affect AMIP robustness through the noise}
\pointlabel{outliers}
Consideration of gross-error robustness encourages users to examine their data
for unusual ``outliers'' in the data; once outliers are removed or their
influence diminished, the problem is considered gross-error robust.  Since
outliers are heuristically associated with heavy-tailed data distributions, one
might expect the effect of outliers to affect AMIP robustness through the shape
variable $\shape$.  However, our analysis of
\secpointref{amip_decomposition}{shape} shows that gross errors actually {\em
reduce} $\shape$ and so render an estimator more robust for a fixed
$\inflscale$. This observation does not imply that gross errors decrease AMIP
sensitivity.  Rather, gross errors increase AMIP sensitivity through the noise
$\inflscale$. And, as we have seen, effects on $\inflscale$ also affect the
computation of standard errors.

%% file: approximation_accuracy.tex
In \secpointref{ols_what_determines}{ols_small_alpha} we argued that our
approximation was accurate in OLS for small $\alpha$. Now we extend that
argument to the general case. In particular, we state sufficient conditions
under which $\thetafunlin(\w)$ provides a good approximation to
$\thetafun(\thetahat(\w), \w)$ for small $\alpha$ uniformly for $\w \in
W_\alpha$.  Our key result, \thmref{theta fun_accuracy}, holds exactly in finite
samples with bounds that are, in principal, computable. Additionally, the
corresponding bounds can also be expected to hold with probability approaching
one as $N \rightarrow \infty$ under standard assumptions.


\subsubsection{Controlling the residual of a Taylor series}

The linear approximation we use in \eqref{taylor_approx} is a Taylor series, so
its accuracy can be controlled by controlling the Taylor series residual.
\citet{giordano:2019:swiss} states conditions under which the first-order Taylor
series approximation to $\thetahat(\w)$ is accurate---precisely when using the
derivative as given in \eqref{dtheta_dw_general}. Under additional smoothness
assumptions on $\thetafun$, we can extend those results to our present
\eqref{taylor_approx}.  Since the Taylor series expansion is expressed in terms
of observable non-asymptotic quantities, the resulting error bounds hold exactly
in finite sample and are, in principle, computable.

We first state assumptions under which the linear approximation is accurate for
the vector $\thetahat(\w)$.

\begin{assu}[(\citet{giordano:2019:swiss}, Assumptions 1-4)]
    \assulabel{ij_assu}
Let $W_\alpha$ be the set of weight vectors with no more than $\lfloor \alpha N
\rfloor$ zeros as given by \eqref{w_alpha_def}.   Assume there exists a compact
domain $\thetadom \subseteq \mathbb{R}^D$ containing $\thetahat(\w)$ for all $\w
\in W_\alpha$, such that
\begin{enumerate}
    \item For all $\theta \in \thetadom$ and all $n$, $\theta \mapsto G(\theta,
    d_n)$ is continuously differentiable with derivative
    \begin{align*}
    \fracat{\partial G(\theta, \d_n)}{\partial \theta^T}{\theta}
    =: H(\theta, \d_n).
    \end{align*}
    \sloppy \item For all $\theta \in \thetadom$, there exists $\cop < \infty$
    such that
    $\sup_{\theta \in \thetadom}\vnorm{\left( 
        \meann H(\theta, d_n)\right)^{-1}
    }_{op} \le \cop$.
    \sloppy \item There exists a constant $\cgh < \infty$ such that
    \begin{align*}
    \sup_{\theta \in \thetadom}
        \max\left\{\meann \vnorm{G(\theta, \d_n)}_2^2,
              \meann \vnorm{H(\theta, \d_n)}_2^2 \right\} \le \cgh^2.
    \end{align*}
    \item There exists a $\Delta_\theta$ and an $\lh < \infty$ such that
    \begin{align*}
    \sup_{\theta: \vnorm{\theta - \thetahat}_2 \le \Delta_\theta}
    \meann \vnorm{H(\theta, \d_n) - H(\thetahat, \d_n)}_2^2 /
         \vnorm{\theta - \thetahat}_2^2
    \le \lh^2.
    \end{align*}
\end{enumerate}
\end{assu}


Roughly speaking, \assuref{ij_assu} states that the estimating equation is
smooth and non-singular, that the sample averages are uniformly bounded, and
that the estimating equation's derivatives are Lipschitz.  Other than the size
of the domain $\thetadom$, \assuref{ij_assu} does not depend on $W_\alpha$, nor
on any asymptotic quantities; it states only (reasonable) assumptions on the
actual problem at hand.

Under \assuref{ij_assu}, we are able to apply Theorem 1 of
\citet{giordano:2019:swiss} for $W_\alpha$ and thereby prove the uniform accuracy
of a linear approximation to $\thetahat(\w)$ for all $\w  \in W_\alpha$.
To extend the accuracy of an approximation of $\thetahat(\w)$ to our
quantity of interest $\thetafun$ naturally requires smoothness assumptions
on $\thetafun$, which we now state.

\begin{assu}\assulabel{thetafun_smooth}
Define the re-scaled weights $\delta_n := \w_n / \sqrt{N}$, and assume that
$\theta, \delta \mapsto \thetafun(\theta, \sqrt{N} \delta)$ has continuous
partial derivatives, that the partial derivatives' $\vnorm{\cdot}_2$-norm
evaluated at $\theta = \thetahat(\onevec)$ and $\w = \onevec$ is bounded by a
finite constant $C_\phi$, and that the partial derivatives are Lipschitz in
$\vnorm{\cdot}_2$ with finite constant $L_\phi$.
\end{assu}

We can now state our main accuracy theorem.

\begin{thm}\thmlabel{thetafun_accuracy}
Let \assuref{ij_assu, thetafun_smooth} hold. For sufficiently small $\alpha$,
there exist constants $C_1$ and $C_2$, defined in terms of quantities given in
\assuref{ij_assu, thetafun_smooth}, such that\footnote{ We note that the rate
$\sqrt{\alpha}$ is determined by a simple but coarse Cauchy-Schwartz bound (see
\lemref{alpha_complexity}).  Tighter bounds may be achievable when the random
variables $\vnorm{G(\theta, \d_n)}_2$ and $\vnorm{H(\theta, \d_n)}_2$ are
uniformly integrable (see, e.g., Section 2.5 of \citet{vaart2000asymptotic}). }
\begin{align}
\sup_{\w \in W_\alpha} \vnorm{\thetafunlin(\w) - \thetafun(\thetahat(\w), \w)}
    \le{}& C_1 \alpha \quad \textrm{and} \quad
\sup_{\w \in W_\alpha} \vnorm{\thetafun(\thetahat(\w), \w) - \thetafunhat}
    \le{} C_2 \sqrt{\alpha}. \eqlabel{thetafun_accuracy}
\end{align}
\end{thm}

When $\alpha$ is small, we expect $\alpha \ll \sqrt{\alpha}$ (for example, when
$\alpha = 0.01$, $\sqrt{\alpha} = 0.1 \gg 0.01$), so \thmref{thetafun_accuracy}
states that the bound in the error of our linear approximation shrinks faster
than the bound in the function itself as $\alpha \rightarrow 0$.
In \appref{tight_bound}, we show that the dependence on $\alpha$ given in
\thmref{thetafun_accuracy} are tight, i.e., that there exist problems for
which the effect size and error scale as $\sqrt{\alpha}$ and $\alpha$,
respectively, as $\alpha \rightarrow 0$.

\Thmref{thetafun_accuracy} is a finite-sample result, applying exactly to the
problem at hand.  All else equal, finite-sample results are preferable to
asymptotic ones. Nevertheless, due to the many loose bounds employed in the
proof, we do not expect the constants to be useful in practice. Additionally,
Theorem 1 of \citet{giordano:2019:swiss} may in theory require $\alpha$ to be
smaller than $1 / N$, resulting in a vacuous statement. Improving these
shortcomings is an important avenue for future work (e.g.
\citet{giordano2019higherorder, wilson:2020:approximate}).  But it is therefore
useful to observe that, when uniform laws of large numbers apply to $\theta
\mapsto \vnorm{G(\theta, \cdot)}_2$ and $\theta \mapsto \vnorm{H(\theta,
\cdot)}_2$, and the limiting functions are also non-singular, bounded, and
Lipschitz, then one can expect \assuref{ij_assu} to hold with high probability
and finite constants as $N \rightarrow \infty$. A precise statement of the
necessary conditions for such asymptotics to apply is given in Lemma 1 of
\citet{giordano:2019:swiss}.

\subsubsection{Limitations of linear approximations}

In every case we examine in our applications in \secref{examples}, we manually
re-run the analysis without the data points in the removal set $\amis{\alpha}$;
in doing so, we find that the change suggested by the approximation is  nearly
always achieved in practice (a notable exception is given and discussed at the
end of \secref{example_microcredit_hierarchical}).  However, linear
approximations are only approximations, and intuition about the potential
weaknesses of linear approximations in general apply to our approximation.  The
crux of \thmref{thetafun_accuracy} is that small $\alpha$ implies that $\w -
\onevec$ is small, thus we can control the error of a linear approximation in
$\w$ evaluated at $\onevec$.  Conversely, one would not expect the approximation
to work well in general for large $\alpha$ and the correspondingly larger $\w -
\onevec$.

As an extreme example, consider when the linear approximation reports that there
is no feasible way to effect a particular change; i.e., when $\aloprop{\Delta} =
\na$ (see \defref{approx_metrics}).  Such a result may seem to imply that, no
matter how many datapoints one removes, the estimator will not change by an
amount $\Delta$, which is often absurd. However, such a result should be taken
to mean that one would have to remove such a large proportion $\alpha$ of
datapoints that the linear approximation on which we are basing the
$\aloprop{\Delta}$ is invalid.  A more accurate interpretation of
$\aloprop{\Delta} = \na$ is that no {\em small} proportion of points can be
removed to produce a change $\Delta$, for if there were such a small proportion,
the linear approximation would have discovered it.

Similarly, linear approximations cannot be expected to work well near the
boundary of parameter spaces.  For example, if the quantity of interest is a
variance, then the true parameter is constrained to be positive, but our linear
approximation is not. It can help to linearize the problem using unconstrained
reparameterizations (e.g., linearly approximating the log variance rather than
variance).  However, as we show in \secref{example_microcredit_hierarchical},
simply transforming to an unconstrained space is still not guaranteed to produce
accurate approximations near the boundary in the original, constrained space.

%% file: related_work.tex
The present work belongs to an extensive ``local robustness'' literature, which
is concerned with measuring robustness using local properties of an estimator
such as series approximations.  In particular, our reliance on the influence
function and its related properties is shared with a great deal of the existing
statistical robustness literature.  Arguably beginning with
\citet{mises1947asymptotic}, the idea of forming series expansions in the space
of data distributions was developed both for the purposes of asymptotic theory
\citep[e.g.][]{jaeckel:1972:infinitesimal,reeds1976thesis,fernholz:1983:mises,vaart:1996:empiricalprocesses},
design of robust estimators \citep[e.g.][chapter
2.4]{hampel1974influence,hampel1986robustbook}, and the detection of
``outliers''
(e.g., \citealt[chapter 2]{belsley:1980:regression};
\citealt{cain:1984:approximatecaseinfluence}; \citealt{cook:1986:assessment}).
Further, the influence function itself is a specific instance of a much broader
idea of differentiating a model with respect to its inputs in order to assess
sensitivity to generic perturbations (e.g. \citealt{cook:1986:assessment} again;
\citealt{diaconis:1986:bayesconsistency};
\citealt{ruggeri:1993:infinitesimalposteriorsensitivity};
\citealt{basu:1996:local}, \citealt{gustafson:2012:localrobustnessbook};
\citealt{giordano2022bnp}). Our work follows in and is deeply indebted to this line
of work.

The general form for the influence function of Z-estimators which we reproduce
in \secref{taylor_series} has been noted many times before in the statistics
literature (e.g., \citealt[chapter 3.4]{hampel1986robustbook};
\citealt{taylor:1993:unifiedapproachtoinfluentialdata}; \citealt[example
20.4]{vaart2000asymptotic}), the machine learning literature (e.g.
\citealt{koh:2017:blackbox}; \citealt{giordano:2019:swiss}), and is of course
simply a consequence of the well-known implicit function theorem
\citep{krantz2012implicit}.  Despite this recognition, there are many examples
of special cases being derived in detail for particular models (e.g.
\citealt{pregibon:1981:logistic}; \citealt{thomas:1989:assessing},
\citealt{hattori:2009:ipcqcasedeletion}; \citealt{shi:2016:gmmcasedeletion}),
suggesting that the simplicity of the general form of the derivative may be
under-appreciated. As we argue in \secref{AMIP}, this general form is
particularly useful to recognize in the age of high-quality automatic
differentiation software.

Our focus on dropping data rather than ``gross errors,'' though not without
precedent, is distinct from much of the robustness literature.  Beginning with
\citet{huber:1964:robustlocation}, much of the statistical robustness literature
has been concerned with the possibility that the model distribution may have
been contaminated with an arbitrarily adversarial distribution or, equivalently,
the observed dataset contains values that can take on arbitrarily misleading
values. In contrast, we focus on dropping asymptotically non-vanishing amounts
of data, which remains a model-agnostic data perturbation while being less
adversarial --- and arguably more reasonable in certain settings, such as
generalization to slightly different populations --- than data that takes on
arbitrarily adversarial values.

Our ``perturbation-inducing proportion'' can be thought of as an example of a
``breakdown point,'' when the latter is defined broadly as ``the proportion of
data which can be changed in some way before something bad happens to the
estimator.''  In the tradition of concern with gross errors, the breakdown point
literature is primarily concerned with the amount of data that can be changed to
an arbitrary degree before an estimator can be changed by an arbitrarily large
amount \citep[chapter 1.2.5]{huber1981robust}.  Our concern, of course, is
different: we only drop data and consider ``something bad'' to be a meaningful
but finite change to a key quantity of interest. Early work such as
\citet{huber:1983:notion} raises the possibility of more generic notions of
breakdown points such as ours. However, as far as the authors are aware, the
present work is the first to pursue our particular notion of breakdown point in
detail.

The concern with gross errors has also led to a large literature which aims to
detect and define ``outliers'' in a context-agnostic way
\citep[e.g.][]{belsley:1980:regression,cook:1982:residualsandinfluenceinregression,cook:1986:assessment,kempthorne:1986:decision,carlin:1991:expectedutilityinfluence}.
Following \citet{cook:1977:detectionofinfluential}, much of this literature
focuses, like us, on the effect of removing datapoints, though typically only on
one or a small number of datapoints.  Furthermore, this line of work evaluates
the effect of dropping datapoints in service of defining a context-agnostic
notion of ``outlier'' rather than focusing, as we do, on a particular decision
using the dataset at hand.

A number of authors in the outlier detection literature consider the removal of
multiple points.  Since their focus is always on identifying a small number of
outliers, they do not consider, as we do, the inferential implications of or the
accuracy of the linear approximation for leaving out a small, fixed proportion
of the data.  For example, \citet[chapter 5]{hadi:2009:sensitivityinregression}
derives straightforward versions of classical ``outlier'' metrics such as Cook's
distance and Andrews-Pregibon statistics for multiple datapoints. \citet[section
2.1]{belsley:1980:regression} discuss ``multiple-row effects'' for linear
regression: motivated by the possibility that groups of points may be
influential collectively but not individually, they propose a stepwise scheme
for finding influential groups of observations based on repeatedly re-fitting
the model, leaving the single most influential point out at each step.
\citet{johnson:1983:predictiveinfluence} considers the effect on a posterior
predictive distribution of the removal of three points out of a set of
twenty-four, which is tractable because of the closed-form solution and
relatively small number of combinations.  \citet{huh:1990:local} observes
briefly that the first-order approximation to leaving out multiple points is the
sum of their influence scores, a fact which they use to produce low-dimensional
visual summaries of effects of groups of observations.
\citet{taylor:1993:unifiedapproachtoinfluentialdata} observes that influence
functions can estimate the effect of leaving out large numbers of datapoints but
consider it not useful, since their primary objective is detecting small numbers
of gross errors.

To the best of the authors' knowledge, our analysis of the effects of leaving
out a non-vanishing proportion of the data, both on the accuracy of the
empirical influence function and on inferential conclusions, is new.

%
%
%

%% file: example_medicaid.tex
In our first experiment, we show that even empirical analyses that display
little classical uncertainty can be sensitive to the removal of less than 1\% of
the sample. We consider the Oregon Medicaid study \citep{finkelstein2012oregon}
and focus on health outcomes. The standard errors of the treatment effects are
small relative to effect size; against a null hypothesis of no effect, most $p$
values are well below 0.01. Yet we find that for most of the results, removing
less than 1\% of the sample can produce a significant result of the opposite
sign to the full-sample analysis. In one case, removing less than 0.05\% of the
sample can change the significance of the result.

\subsubsection{Background and replication}
First we provide some context for the analysis and results of
\citet{finkelstein2012oregon}. In early 2008, the state of Oregon opened a
waiting list for new enrollments in its Medicaid program for low-income adults.
Oregon officials then drew names by lottery from the 90,000 people who signed
up, and those who won the lottery could sign up for Medicaid along with any of
their household members. This setup created a randomization into treatment and
control groups at the household level. The \citet{finkelstein2012oregon} study
measures outcomes one year after the treatment group received Medicaid. About
25\% of the treatment group did indeed have Medicaid coverage by the end of the
trial. The main analysis investigates treatment assignment as treatment itself
(``intent to treat'' or ITT analysis) and uses treatment assignment as an
instrumental variable for take-up of insurance coverage (``local average
treatment effect'' or LATE analysis).

We focus on the health outcomes of winning the Medicaid lottery, which appear in
Panel B from Table 9 of \citet{finkelstein2012oregon}. Each of these $J$
outcomes is denoted by $y_{ihj}$ for individual $i$ in household $h$ for outcome
type $j$. The data sample to which we have access consists of survey responders
($N = 23{,}741$); some responders are from the same household.  The variable
$\texttt{LOTTERY}_h$ equals one if household $h$ won the Medicaid lottery, and
zero otherwise. All regressions use a set of covariates $X_{ih}$ comprised of
household size fixed effects, survey wave fixed effects, and the interaction
between the two. All regressions also use a set of demographic and economic
covariates $V_{ih}$. To infer the ITT effects of winning the Medicaid lottery,
the authors estimate the following model via OLS: \begin{align*}
y_{ihj} = \beta_0 + \beta_1 \texttt{LOTTERY}_h + \beta_2 X_{ih} +
    \beta_3 V_{ih} + \epsilon_{ihj}.
\end{align*}

To infer the LATE of taking up Medicaid on compliers, the authors employ an
Instrumental Variables (IV) strategy using the lottery as an instrument for
having Medicaid insurance. All standard errors are clustered on the household,
and all regressions are weighted using survey weights defined by the variable
\texttt{weight\_12m}. We have access to the following seven outcome variables,
presented in Panel B of Table 9 of the original paper (as well as our tables
below) in the following order: a binary indicator of a self-reported measure of
health being good or very good or excellent (not fair or poor), a binary
indicator of self-reported health not being poor, a binary indicator of health
being about the same or improving over the last six months, the number of days
of good physical health in the past 30 days, the number of days on which poor
physical or mental health did not impair usual activities, the number of days
mental health was good in the past 30 days, and an indicator of not being
depressed in last two weeks. We replicate Panel B of Table 9 of
\citet{finkelstein2012oregon} exactly, both for the ITT effect ($\hat{\beta}_1$)
for the entire population and for the LATE on compliers ($\hat{\pi}_1$). Both
analyses show strong evidence for positive effects on all health measures, with
most $p$ values well below 0.01.

\subsubsection{AMIP Sensitivity Results}

\OHIEResultsTable{}

For each health outcome in Panel B from Table 9 of
\citet{finkelstein2012oregon}, we compute the AMIP to assess how many data
points one needs to remove to change the sign of the treatment effect, the
significance of the treatment effect, or produce a significant result of the
opposite sign. The sensitivity of the LATE analysis is shown in
\tableref{ohie_profit_results_iv} and the sensitivity of the ITT analysis is
shown in \tableref{ohie_profit_results_reg}. In both cases we use exactly the
models from the original paper, with all fixed effects and controls included and
with clustering at the household level. For most outcomes, for both the LATE and
ITT analysis, the sign of the treatment effect can be changed by removing around
0.5\% of the data, or approximately 100 data points in a sample of approximately
22,000. The most robust outcome, ``Health being better than fair'' (``Health
genflip 12m''), requires the removal of a little over 1\% of the sample to
change the sign. Across the various outcomes, we can drop even less of the
sample to change the results from significant to non-significant. In some cases,
we need remove only 10 or 20 data points to effect a change in significance.
Finally, for most outcomes, we can remove less than 1\% of the data to produce a
significant result of the opposite sign. The only two exceptions, ``Health
genflip 12m'' and ``Health change flip 12m'', require the removal of slightly
more than 1\% to generate a significant result with the opposite sign.

We check the performance of the approximation for each analysis by re-running
the model after manually removing the data points in the Approximate Most
Influential Set. The result of this procedure is shown in the ``Refit Estimate''
column of \tableref{ohie_profit_results_iv, ohie_profit_results_reg}. For almost
every result in each table, our approximate metric reliably uncovers
combinations of data points that do deliver the claimed changes. As we discuss
in \secref{exact_lower_bound}, the changes recorded in the ``Refit Estimate''
column of \tableref{ohie_profit_results_iv, ohie_profit_results_reg} form a
lower bound on the true worst-case finite-sample sensitivity.

By comparing \tableref{ohie_profit_results_iv} with
\tableref{ohie_profit_results_reg}, we see that the ITT results, estimated via
OLS, are not notably more AMIP-robust than the LATE results, which are estimated
via IV. This may seem at first counterintuitive based on a heuristic belief that
IV is in some sense a less ``robust'' analysis than OLS in finite sample: for
example, recent authors, including \citet{young2019consistency}, have suggested
that the uncertainty intervals for IV may be more poorly calibrated in finite
samples than the intervals for OLS.  However, as we discuss in \secref{why}, the
quality of being ``robust'' in the sense of a finite-sample estimator providing
a good approximation to an asymptotic quantity is simply unrelated to AMIP
robustness.  Neither the size of the AMIP itself nor the accuracy of the AMIP
approximation depends on asymptotic arguments (see, e.g.,
\secpointref{amip_robustness_breakdown}{snr_drives_amip} and the discussion of
\thmref{thetafun_accuracy}).  The AMIP measures the sensitivity to data ablation
of a particular procedure on a particular dataset and is indifferent to the
fidelity of the chosen quantity of interest to some asymptotic limit.  For this
reason, a procedure such as IV may be ``non-robust'' in the sense of having poor
coverage in finite sample (as reported by \citet{young2019consistency}) and yet
be AMIP-robust, or vice versa. The two notions of ``robustness'' are simply
different.


%% file: example_transfers.tex
We next show that an empirical analysis can still be AMIP-non-robust even after
outliers are removed. To that end, we apply our techniques to examine the
robustness of the main analysis from \citet{angelucci2009indirect}, one of the
flagship studies showing the impact of cash transfers on ineligible
(``non-poor'') households in the same villages, also known as ``spillover
effects.'' The authors trimmed the consumption outcome for the non-poor
households due to concerns about the influence of the largest values. Yet while
the analysis on the poor households is quite robust, the analysis on the
non-poor households---whom the trimming protocol actually affects---is much more
sensitive.

\subsubsection{Background and replication}

\citet{angelucci2009indirect} employ a randomized controlled trial to study the
impact of Progresa, a social program giving cash gifts to eligible poor
households in Mexico. The randomization occurs at the village level. So one can
estimate both a main effect on the poor households selected to receive Progresa
and also the impact on the non-eligible ``non-poor'' households located in the
same villages as Progresa-receiving poor households.

The main results of the paper show that there are strong positive impacts of
Progresa on total household consumption measured as an index both for eligible
poor households and for the non-eligible households; see Table 1 of
\citet{angelucci2009indirect}. The variable $\texttt{C\_ind}_{it}$ denotes total
household consumption for household $i$ in time period $t$. Values of
$\texttt{C\_ind}_{it}$ above 10,000 are removed; such households are, by
definition, non-poor. The authors study three different time periods separately
to detect any change in the impact between the short and long term. They
condition on a large set of variables (a household poverty index, land size,
head of household gender, age, whether the household speaks an indigenous
language, and literacy; at the locality level, a poverty index, and the number
of households) to help ensure a fair comparison between households in the
treatment and control villages. In this case these controls are important; the
effects on the ``non-poor'' households are significant at the 5\% level when the
controls are included, but they are only significant at the 10\% level in a
simple regression on a dummy for treatment status.

The full data for the paper is available on the website of the \emph{American
Economic Review} thanks to the open-data policies of the journal and the
authors. We can successfully replicate the results of this analysis with the
controls and without, and we proceed with the controls in our present analysis
in accordance with the original authors' preferred specification. We consider
the time periods indexed as $t=8,9,10$ in the dataset provided, though we note
that the authors do not rely on the results at $t = 8$ as the roll-out was still
ongoing. We employ $K$ control variables, where $X_{itk}$ is the $k$-th variable
for household $i$ in period $t$. Then we run the following regression:
\begin{align*}
\texttt{C\_ind}_{it} = \beta_0 + \beta_1 \texttt{treat}_{poor,i} + \beta_2
\texttt{treat}_{nonpoor,i} + \sum_{k = 1}^K \beta_{2+k}X_{itk} + \epsilon_{it}.
\end{align*}
Here, $\texttt{treat}_{poor,i}$ refers to an interaction between the treatment
indicator and an indicator for being a poor household; correspondingly,
$\texttt{treat}_{nonpoor,i}$ is an interaction between the treatment indicator
and an indicator for being a non-poor household.  We are able to exactly
replicate the results of Table 1 of \citet{angelucci2009indirect}, which
exhibits positive effects of cash transfers.

\subsubsection{AMIP Sensitivity Results}

\CashTransfersResultsTable{}
We apply our methodology to assess how many data points one need remove to
change the sign, the significance, or to generate a significant result of the
opposite sign to that found in the full sample. We focus on the latter two time
periods, as households had received only partial transfers in the first time
period, but we show all three in order to replicate Table 1 from the original
paper. \tableref{cash_transfers_re_run_table} shows our results. Focusing on
periods 9 and 10, we find that the inferences on the direct effects on the poor
households are quite robust, but the inferences on the indirect effects are less
so. For the analysis of the poor, one typically needs to remove much more than
1\% of the sample to change conclusions. For the analysis of the non-poor, we
can remove less than 0.1\% of the data to change conclusions. In fact, we can
remove only 3 data points to change the significance status for both $t = 9$ and
$t = 10$.

We again check the quality of our approximation. The ``Refit Estimate'' column
in \tableref{cash_transfers_re_run_table} shows the results of manually
re-running each analysis after removing the implicated data points. In most
cases the AMIP correctly identifies a combination of data points that can make
the claimed changes to the conclusions of the study. Although there are a few
cases where re-running the analysis fails to produce the predicted statistically
significant sign change, the observed changes are still large enough to be of
practical interest.  Furthermore, it is likely that the removal of a few
additional points would in fact produce the desired statistically significant
sign reversals.

Finally, we note that these results constitute an illustration of how gross
error robustness is distinct from AMIP robustness (see
\secpointref{influence_function_for_real}{gross_errors}). Recall that
\citet{angelucci2009indirect} removed (non-poor) datapoints for which
consumption was greater than 10,000. By removing outliers of the consumption
variable in this way, the authors of this study made what is typically
considered a conservative choice in view of classical robustness concerns about
gross error sensitivity. Yet, as we have shown in
\tableref{cash_transfers_re_run_table}, qualitative conclusions concerning the
non-poor households remain non-robust to the removal of a small number of
datapoints, which demonstrates empirically that one cannot necessarily make an
analysis AMIP-robust by simply trimming outliers.  Indeed, as we showed above in
\secref{influence_function_ols}, even perfectly specified OLS regressions with
no aberrant data points can be AMIP-non-robust if the signal to noise ratio is
too low.

%% file: example_microcredit_linear.tex
We now show that even a simple 2-parameter linear model that performs a
comparison of means between the treatment and control group of a randomized
trial can be highly sensitive. To that end, we consider the analysis of seven
randomized controlled trials of expanding access to microcredit, first
aggregated in \citet{meager2019understanding}.  In
\secref{example_microcredit_hierarchical} below, we will consider a more
complicated Bayesian hierarchical model on the same data.

\subsubsection{Background}
Each of the seven microcredit studies was conducted in a different country, and
each study selected certain communities to randomly receive greater access to
microcredit. Researchers either built a branch, or combined building a branch
with some active outreach, or randomly selected borrowers among those who
applied. The selected studies are:
\citet{angelucci2015microcredit}, \citet{attanasio2015impacts},
\citet{augsburg2015impacts}, \citet{banerjee2015miracle},
\citet{crepon2015estimating}, \citet{karlan2011microcredit}, and
\citet{tarozzi2015impacts}.
Six of these studies were published in a special issue of the \emph{American
Economics Journal: Applied Economics} on microcredit. All seven studies together
are commonly considered to represent the most solid evidence base for
understanding the impact of microcredit.

We follow the original studies and \citet{meager2019understanding} in analyzing
the impact of access to microcredit as the treatment of interest. The studies
range in their sample sizes from around 1,000 households in Mongolia
\citep{attanasio2015impacts} to around 16,500 households in Mexico
\citep{angelucci2015microcredit}. We first focus on the headline results on
household business profit regressed on an intercept and a binary variable
indicating whether a household was allocated to the treatment group or to the
control group. For household $i$ in site $k$, let $Y_{ik}$ denote the profit
measured, and let $T_{ik}$ denote the treatment status. We estimate the
following model via OLS:
\begin{equation}\eqlabel{mc_regression}
	Y_{ik} = \beta_0 + \beta T_{ik} + \epsilon_{ik}.
\end{equation}

This regression model compares the means in the treatment and control groups and
estimates the difference as $\hat{\beta}$. We follow
\citet{meager2019understanding} in omitting the control variables or fixed
effects from the regressions in order to examine the robustness of this
fundamental procedure. But in principle this omission should make no difference
to the estimate $\hat{\beta}$, and indeed it does not
\citep{meager2019understanding}.\footnote{The omission may in principle make a
difference to the inference on $\beta$ by affecting the standard errors.
However, it turns out that in these studies the additional covariates make very
little difference to the standard errors. We also do not cluster the standard
errors at the community level for the same reason; the results are not
substantially changed. Running the regression above in each of the seven studies
delivers almost identical results to the preferred specification, as it should
if intra-cluster correlations are weak and covariates are not strongly predictive
of household profit.}

\subsubsection{AMIP sensitivity results}
\MicrocreditProfitResultsTable{}
\MicrocreditTemptationResultsTable{}
The sensitivity results for the linear regression of profit on microcredit
access appear in \tableref{mc_profit_results}. In all cases, by removing less
than 1\% of the data points can change either the sign or the significance. In
three of the studies, one can drop less than 1\% of the data points to generate
a result of the opposite sign that would be deemed significant at the 5\% level.
Mexico, the largest study, is the most sensitive: a single data point among the
16,561 households in Mexico determines the sign (as also discussed above in
\secref{linear_regression}). To produce a statistically significant result of
the opposite sign---that is, to turn Mexico's noisy negative result into a
``strong'' positive result---one need remove only 15 data points, less than
0.1\% of the sample. Mongolia, the smallest study in terms of sample size, is
among the most robust in terms of sign changes; it takes 2\% of the sample to
change the sign. Producing a significant result of the opposite sign also
requires more than 1\% removal in the Philippines, Bosnia, Ethiopia, and
Mongolia---whereas Mexico, India, and Morocco are more sensitive. We check the
performance of our approximation by manually re-running the analysis with the
data removed; the ``Refit Estimate'' column shows that the claimed reversal is
always achieved in practice for these analyses.

By comparing the results of the present section with those of
\secref{example_medicaid,example_transfers}, we can confirm the conclusion of
\secpointref{amip_robustness_breakdown}{amip_is_not_se} that standard errors
are, in general, distinct from AMIP sensitivity.  Despite the fact that original
estimates of \tableref{mc_profit_results} are statistically insignificant, some
of these non-significant results are more AMIP-robust than some of the
significant results in the Cash Transfers and Oregon Medicaid examples; consider
the ``Significant sign change'' result in the Philippines study, for example.

We next demonstrate that the AMIP sensitivity observed in
\tableref{mc_profit_results} cannot simply be ascribed to statistical
insignificance. To do so, we consider a different outcome with smaller variability
and show that it reveals a similar sensitivity to the profit outcome.
The variable we now consider is household consumption spending on temptation goods such
as alcohol, chocolate, and cigarettes, since the effect of microcredit on
temptation spending was estimated by \citet{meager2019understanding} with the
greatest precision of all six considered outcome variables.
\Tableref{mc_temptation_results} shows the results of applying the AMIP to  the
same regression given in \eqref{mc_regression}, but with temptation spending as the outcome. While
somewhat more robust than the profit analyses, the difference in the approximate
removal proportions in \tableref{mc_temptation_results} is not large.

Finally, one might be tempted to ascribe the AMIP-non-robust results in
\tableref{mc_profit_results} to outliers resulting from the heavy tails of the
household profit variable (a phenomenon well-documented by
\citet{meager2020aggregating}). However, as we discuss in
\secpointref{influence_function_for_real}{gross_errors} above, gross error
robustness is qualitatively distinct from AMIP robustness (see also the
discussion of outlier trimming at the end of \secref{example_transfers}).
Indeed, the more complex hierarchical model of the next section,
\secref{example_microcredit_hierarchical}, was designed precisely to accommodate
the heavy tail of the household profit variable, and yet---as we will
show---still exhibits a high degree of AMIP-sensitivity.

%% file: example_microcredit_hierarchical.tex

In this section, we investigate a Bayesian hierarchical model, both
demonstrating that even Bayesian analyses can exhibit considerable AMIP
sensitivity, and showing an example of a parameter of interest for which our
linear approximation performs badly. We specifically focus on a variational
Bayes approximation to the tailored mixture model from
\citet{meager2020aggregating}.  One might hope that any of the following aspects
of the more complicated model might alleviate AMIP sensitivity: the use of
hierarchical Bayesian evidence aggregation, the regularization from
incorporation of priors, or the somewhat more realistic data-generating process
captured in this specific tailored likelihood. Indeed, the approach of
\citet{meager2020aggregating} was specifically motivated by the desire to
capture important features of the data-generating process such as heavier tails.
On the contrary, we find that the average  estimated effects of microcredit
remain sensitive according to the AMIP, as we did in the simpler models of
\secref{example_microcredit_linear}.  We also find that the linear approximation
that underlies the AMIP performs poorly when attempting to decrease a particular
hypervariance parameter, providing a concrete example of the limitations of our
methodology, particularly for parameters near the boundary of the set of their
allowable values.

\subsubsection{Background}

Following \citet{meager2020aggregating}, we fit a hierarchical model (hereafter
referred to as the ``microcredit model'') to all the data from the seven
microcredit RCTs. We model each outcome using a spike at zero and two lognormal
tail distributions, one for the positive realizations of profit and one for the
negative realizations. Within the model, microcredit can affect the proportion
of data assigned to each of these three components as well as affecting the
location and scale of the lognormal tails. There is a hierarchical shrinkage
element to the model for each parameter. The hypervariances of the treatment
effects are of particular interest because these capture heterogeneity in
effects across studies and offer information about the transportability of
results across settings.

The models in the original paper were fit via Hamiltonian Monte Carlo (HMC) with
the software package Stan \citep{carpenter2017stan}. It is possible to compute
the Approximate Maximum Influence Perturbation for HMC, or for any Markov Chain
Monte Carlo method, using the tools of Bayesian local robustness
\citep{gustafson2000local, giordano2018covariances}, but the sensitivity of
simulation-based estimators is beyond the scope of this paper. However, there
are ways to estimate Bayesian posteriors via Z-estimators, such as with
Variational Bayes (VB) techniques \citep{blei2016variational}.\footnote{The
Laplace approximation can also be expressed as a Z-estimator.} Specifically, we
fit the microcredit model using a variant of Automatic Differentiation
Variational Inference (ADVI) described in \citet[Section
5.2]{giordano2018covariances} (see also the original ADVI paper,
\citet{kucukelbir2017advi}).  Since the posterior uncertainty estimates of
vanilla ADVI are notoriously inaccurate, we estimated posterior uncertainty
using linear response covariances, again following \citet[Section
5.2]{giordano2018covariances}.\footnote{When forming the Approximate Most
Influential Set, we approximated the sensitivity only of the posterior means to
data removal; the linear response covariances were considered fixed.  However,
when we report the results of re-fitting the model, we did re-calculate the
linear response covariances at the new variational optimum.}  We verified that
the posterior means and covariance estimates produced by our variational
procedure and the corresponding estimates from running HMC with Stan were within
reasonable agreement relative to the posterior standard deviation.

\subsubsection{AMIP Sensitivity Results}

\MicrocreditMixtureResultsTable{}

\MicrocreditMixtureSdResultsTable{}

We first consider the effect of microcredit on the location parameter of the
positive and negative tails of profit, given respectively by the parameters
$\tau_{+}$ and $\tau_{-}$.  Roughly speaking, $\tau_{+}$ and $\tau_{-}$ are both
estimating the effect of microcredit averaged across all of the seven countries
analyzed in \secref{example_microcredit_linear}.  Our point estimates for
$\tau_{+}$ and $\tau_{-}$ are given by their respective VB posterior means.  We
used the linear response covariance estimates to form a 95\% posterior credible
interval in place of confidence intervals, and consider a change ``significant''
if the posterior credible interval does not contain zero.

\tableref{mcmix_re_run_table} shows the sensitivity of inference concerning
$\tau_{+}$ and $\tau_{-}$.  We see that the microcredit model's estimates of the
average effectiveness of microcredit remain highly sensitive to the removal of
small percentages of the sample, despite being derived from a model that
accounts for non-Gaussian data shape and is regularized by the priors. This
sensitivity shows that Bayesian aggregation procedures do not necessarily
produce AMIP-robust estimates.

We next examine the sensitivity of the hypervariances, which measure the
variability of the effect of microcredit on these tails from country to country.  Specifically,
the parameters $\sigma_{\tau_{+}}^2$ and $\sigma_{\tau_{-}}^2$ represent the
between-country variances of the effect of microcredit on positive and negative
profit outcomes, respectively. The $\sigma$ parameter can be thought of as the
scale parameter analogue of the corresponding location parameter $\tau$ from
\tableref{mcmix_re_run_table}. The hypervariances are of particular practical
interest because they quantify how variable the effect of microcredit might
be; small values of the hypervariance imply that all countries respond
similarly to microcredit, whereas large values imply that one should not
necessarily extrapolate the efficacy of microcredit from one country to another.

In order to avoid the possibility of extrapolating to negative variances, we form
a linear approximation to our variational Bayes estimates of the posterior mean
of $\log \sigma$.  Since $\log \sigma$ is a scale parameter measuring the
variability from country to country of the effect of microcredit, its sign is
not particularly meaningful, nor is it particularly interesting to ask whether
its posterior credible interval contains zero.  Rather, we are interested in the
magnitude of $\log \sigma$. So, to investigate robustness, we use the AMIP to
check the approximate maximum change achievable in either direction (increasing
or decreasing the magnitude of $\log \sigma$) by removing 0.5\% of the sample,
about the same fraction of the data as could generate a ``significant'' sign
change for the $\tau_{\pm}$ parameters.

The results for the hypervariances, given in \tableref{mcmix_sd_re_run_table},
represent a useful demonstration of the limitations of our linear approximation.
We are able to find sets of datapoints which, when dropped, produce {\em
increases} in the hypervariances, though the our linear approximation is not
nearly as accurate as in the rest of our results above. When we attempted to
drop points in order to decrease the hypervariances, however, the linear
approximation failed utterly; the Approximate Most Influential Set designed to
produce a decrease in the hypervariances instead produced a large {\em
increase} upon refitting.

Given that the hypervariances are constrained to be positive, our failure to
produce large decreases may not be surprising. Note that the hypervariances'
posterior expectations began very small, and that decreasing them pushes the
posterior of the hypervariances closer to the boundary of the admissible space.
Though the log variance may in principle take arbitrarily negative values, it
nevertheless appears that the model exhibits strongly non-linear dependence on
the data weights for very small variances. Designing useful diagnostics for
detecting and explaining such deviations from nonlinearity in complex models is
an interesting avenue for future work. In the meantime,
\tableref{mcmix_sd_re_run_table} shows the importance, when possible, of
checking the accuracy of the AMIP predictions by refitting the model, and of
exercising caution when using the AMIP approximation near the boundary of the
parameter space.

%% file: conclusion.tex
For our research conclusions to safely inform economic policy decisions, we need
additional tools to quantify uncertainty beyond standard errors. There are many
ways of quantifying the dependence between the finite-sample realization of the
data and the conclusions of statistical inference. This dependence has become
synonymous with standard errors in frequentist statistics, but the notions are
equivalent only under a certain paradigm that considers a hypothetical perfect
random resampling exercise for the purpose of evaluating a specific parameter
within a given model. This hypothetical may not capture all the data sensitivity
relevant to applied social science.

Many key ideas of 20th century statistics have their origins in the context of
randomized agricultural trials, where the difference in yield across multiple
fields is well-modeled by independent sampling variation. Contrast this setting
with trials of economic interventions to alleviate poverty, where randomly
sampling individuals or communities is a challenge and interventions may be
applied across very different contexts. In such cases, statistical models are
often intended to provide tractable and interpretable summaries or proxies of
the general impact of interventions.  As such methods often average information
across individuals in ways that may not always reflect broader policy interests,
it seems essential to interrogate the sensitivity of our conclusions to
departures from the hypothetical thought experiment.\footnote{In agricultural
trials, total yields are the true quantity of interest; for microcredit trials,
the average treatment effect is but a convenient summary. If the average profit
were to increase slightly through one individual becoming wealthy while leaving
all others destitute, one could consider the intervention a failure. By
contrast, if a single plant produced an entire harvest's worth of corn, the
outcome would still be desirable, if strange.}

In this paper, we have offered one alternative way of conceiving of and
quantifying the dependence of empirical results on the sample data, beyond
standard errors. Sensitivity of conclusions to data removal under our metric
does not necessarily imply a problem with the sample. But the goal of inference
is not to learn about the sample, but to learn about the population. If
minor alterations to the sample can generate major changes in the inference, and
we know that the environment in which we do economics is changing all the time,
we ought to be less confident that we have learned something fundamental about
the world we seek to understand, for which we ultimately seek to
make policy. This does not imply that the original analysis is invalid
according to classical sampling theory, and we do not recommend that researchers
abandon the original full-sample results even if they are sensitive according
to our metric. However, reporting our metrics alongside standard errors would
improve our ability to understand and interpret the findings of a given
analysis.

Since AMIP analysis always indicates which data points have high (approximate)
influence, our methods allow researchers not only the chance to check that the
approximation worked on their own sample, but to understand what---if
anything---makes these data points special. Investigating influential points may
provide insight into the way in which a given inferential procedure is using the
finite-sample information to generate claims about the population parameters. In
addition, in cases when this sensitivity is undesirable, it may be fruitful to
develop new statistical methods to ameliorate it.

%% file: appendix_proofs.tex
\begin{lem}\lemlabel{normalized_sum_bound}
\sloppy
Let $\chi_1, \ldots, \chi_N$ be real-valued scalars with $\meann \chi_n^2 = 1$.
Then $\max_{\w \in W_\alpha} \meann \abs{\w_n - 1}\chi_n \le \sqrt{\alpha}$.
\begin{proof}
Without loss of generality, let the $\chi_n$ be unique (if they are not,
add an arbitrarily small amount of jitter to break ties), and let $q_{1 -
\alpha}$ denote their $\lceil (1 - \alpha) N \rceil$-th largest value.
The maximum $\max_{\w \in W_\alpha} \meann \abs{\w_n - 1} \chi_n$ is
achieved at $\w$ which sets to zero the weights of all $\{n : \chi_n \ge
q_{1-\alpha}\}$, so
\begin{align*}
\max_{\w \in W_\alpha} \meann \abs{\w_n - 1}\chi_n =
\meann \ind{\chi_n \ge q_{1 - \alpha}} \chi_n.
\end{align*}
Let $\hat{F}_{\chi}$ denote the empirical distribution on $\chi_n$
conditional on the data $\d_n$, and note that $q_{1 - \alpha}$ is fixed
in $\hat{F}_{\chi}$.  Applying Cauchy-Schwartz to the preceding display
with the distribution $\hat{F}_{\chi}$ gives
\begin{align*}
%
\meann \ind{\chi_n \ge q_{1 - \alpha}} \chi_n
\le \sqrt{\meann \ind{\chi_n \ge q_{1 - \alpha}}^2}
\sqrt{\meann \chi_n^2} =
\sqrt{\frac{\lfloor N \alpha \rfloor}{N}}
\le \sqrt{\alpha},
\end{align*}
since $\meann \chi_n^2 = 1$ and at most $\lfloor \alpha N \rfloor$
points are greater than $q_{1-\alpha}$.
\end{proof}
\end{lem}


The following lemma shows that \assuref{ij_assu} satisfies Condition 1 of
\citet{giordano:2019:swiss}.

\begin{lem}\lemlabel{alpha_complexity}
Let $W_\alpha^* := \{ \onevec + t (\w - \onevec): \w \in W_\alpha, t \in [0, 1] \}$
Under \assuref{ij_assu},
\begin{align*}
\max_{\w \in W_\alpha^*}
\sup_{\theta \in \thetadom}
    \vnorm{\meann (\w_n - 1) G(\theta, \d_n)}_1   \le&
                \sqrt{D} \cgh \sqrt{\alpha} \quad\textrm{and}\\
\max_{\w \in W_\alpha^*}
\sup_{\theta \in \thetadom}
    \vnorm{\meann (\w_n - 1) H(\theta, \d_n)}_1   \le&
                \sqrt{D} \cgh \sqrt{\alpha}.
\end{align*}
\begin{proof}
We prove the result for $G(\theta, \d_n)$; the proof for $H(\theta, \d_n)$
follows analogously.  By the triangle inequality and the relationship between
$\vnorm{\cdot}_2$ and $\vnorm{\cdot}_1$,
\begin{align*}
\vnorm{\meann (\w_n - 1) G(\theta, \d_n)}_1 \le
\sqrt{D} \cgh \meann \abs{\w_n - 1} \frac{\vnorm{G(\theta, \d_n)}_2}{\cgh}.
\end{align*}
Apply \lemref{normalized_sum_bound} with $\chi_n := \frac{\vnorm{G(\theta,
\d_n)}_2}{\cgh}$ to control the maximum of the sum over $W_\alpha$.  Finally,
the results extends to $W_\alpha^*$ since
\begin{align*}
\max_{\w \in W_\alpha^*} \meann \abs{\w_n - 1}\chi_n
=&
\max_{t \in [0,1]} \max_{\w \in W_\alpha}
    \meann \abs{t (\w_n - 1)} \chi_n =
\max_{\w \in W_\alpha} \meann \abs{(\w_n - 1)} \chi_n.
\end{align*}
\end{proof}
\end{lem}


We need the following lemma to extend the result of \citet{giordano:2019:swiss},
Theorem 1 to smooth functions.

\begin{lem}\lemlabel{derivative_smooth}
Let \assuref{ij_assu, thetafun_smooth} hold. For sufficiently small $\alpha$,
there exists a constant $C_b < \infty$ such that, for any $a \in \mathbb{R}^N$,
\begin{align*}
\max_{\w \in W_\alpha^*}
\vnorm{\left(
    \fracat{d \thetahat(\w)}{ d\w^T}{\w} -
    \fracat{d \thetahat(\w)}{ d\w^T}{\onevec}
    \right) a }_2
\le C_b \frac{\vnorm{a}_2}{\sqrt{N}} \sqrt{\alpha}.
\end{align*}
\begin{proof}
As in the proof of \thmref{thetafun_accuracy}, for the remainder of the proof
assume that $\alpha \le \frac{\Delta^2}{D\cgh^2}$, and observe that Assumptions
1-5 and Condition 1 of \citet{giordano:2019:swiss} are satisfied.
For the duration of this proof, define the shorthand notation
\begin{align*}
H(\w) := \meann \w_n H(\thetahat(\w), d_n)
\quad\textrm{and}\quad
G(\w) := \meann a_n G(\thetahat(\w), d_n).
\end{align*}
Then, by the indicated results from \citet{giordano:2019:swiss},
\begin{align*}
\MoveEqLeft
\vnorm{\left(
    \fracat{d \thetahat(\w)}{ d\w^T}{\w} -
    \fracat{d \thetahat(\w)}{ d\w^T}{\onevec}
    \right) a }_2
\\={}&
\vnorm{-H(\w)^{-1} G(\w) + H(\onevec)^{-1} G(\onevec)}_2
    \quad \textrm{(Proposition 4)}
\\\le{}&
\vnorm{-(H(\w)^{-1} - H(\onevec)^{-1} ) G(\w)}_2 +
\vnorm{ H(\onevec)^{-1} ( G(\onevec) - G(\w))}_2
\\\le{}&
\vnorm{-(H(\w)^{-1} - H(\onevec)^{-1} ) G(\w)}_2 + \cop \delta.
    \quad \textrm{(Condition 1, Assumption 2)}
\end{align*}
Then,
\begin{align*}
\MoveEqLeft
\vnorm{(H(\w)^{-1} - H(\onevec)^{-1} ) G(\w)}_2
\\={}&
\vnorm{ H(\w)^{-1} \left(H(\onevec) -  H(\w)\right) H(\onevec)^{-1} G(\w)}_2
\\\le{}&
2\cop^2 \vnorm{\left(H(\onevec) -  H(\w)\right) G(\w)}_2
\quad\textrm{(Assumption 2, Lemma 6)}
\\\le{}&
2\cop^2 \sqrt{D} (1 + D C_w L_h \cop) \delta \vnorm{G(\w)}_2
\quad\textrm{(Lemma 5, Matrix norms)}
\\={}&
2\cop^2 \sqrt{D} (1 + D C_w L_h \cop) \delta
     \vnorm{\meann a_n G(\thetahat(\w), d_n)}_2
\\\le{}&
2\cop^2 \sqrt{D} (1 + D C_w L_h \cop) \delta
     \cgh \frac{\vnorm{a}_2}{\sqrt{N}}.
\quad\textrm{(Assumption 3, Cauchy-Schwartz)}
\end{align*}
Combining, and using our \lemref{alpha_complexity} to give $\delta = \sqrt{D}
\cgh \sqrt{\alpha}$, gives the desired result.
\end{proof}
\end{lem}


\textbf{Proof of \thmref{thetafun_accuracy}.}
\sloppy For the duration of the proof, define the linear approximation
$\thetalin(\w) := \thetahat + \fracat{\dee\thetahat(\w)}{\dee\w^T}{\onevec}(\w -
\onevec)$. \Assuref{ij_assu} is equivalent to Assumptions 1-4 of
\citet{giordano:2019:swiss}, and \lemref{alpha_complexity} satisfies Condition 1
of \citet{giordano:2019:swiss} with $\delta = \sqrt{D} \cgh \sqrt{\alpha}$.
Assumption 5 of \citet{giordano:2019:swiss} is satisfied for $W_\alpha$ with
$C_w = 1$. Define, as in \citet{giordano:2019:swiss}, $\cij := 1 + D \lh \cop$
and $\Delta := \min\left\{\Delta_\theta \cop^{-1}, \frac{1}{2}\cop^{-1}\cij^{-1}
\right\}$, So Lemma~3 and Theorem~1 of \citet{giordano:2019:swiss} give,
respectively, that
\begin{align}
\max_{\w \in W_\alpha^*}
    \vnorm{\thetahat(\w) - \thetahat}_2 \le{}& \cop \sqrt{D} \cgh \sqrt{\alpha}
\quad \textrm{and}  \eqlabel{theta_diff_bound} \\
\alpha \le
\frac{\Delta^2}{D\cgh^2} \quad\Rightarrow\quad
\max_{\w \in W_\alpha^*}\vnorm{\thetalin(\w) - \thetahat(\w)}_2
    \le{}& 2 \cop^2 \cij D \cgh^2 \alpha. \eqlabel{theta_accuracy_bound}
\end{align}
For the remainder of the proof assume that $\alpha \le
\frac{\Delta^2}{D\cgh^2}$ so that \eqref{theta_accuracy_bound} applies.

For any $\w \in W_\alpha$, define $\omega(t) := \onevec + t (\w - \onevec)  \in
W_\alpha^*$. By the fundamental theorem of calculus,
\begin{align}\eqlabel{thetafun_integral}
\thetafun(\thetahat(\w), \w) - \thetafunhat = \int_0^1
    \fracat{\dee \thetafun(\omega(t))}{\dee t}{t}dt
=
\int_0^1
    \left( \fracat{\dee \thetafun(\omega(t))}{\dee t}{t} -
           \fracat{\dee \thetafun(\omega(t))}{\dee t}{1}\right)\dee t +
    \fracat{\dee \thetafun(\omega(t))}{\dee t}{1}.
\end{align}
where, by the chain rule,
\begin{align*}
\fracat{\dee \thetafun(\omega(t)))}{\dee t}{t}
={}&
\fracat{\partial \thetafun(\theta, \omega(t))}
       {\partial \theta^T}{\thetahat(\omega(t))}
\fracat{\dee \thetahat(\w)}{\dee \w^T}{\omega(t)} (\w - \onevec) +
\fracat{\partial \thetafun(\thetahat(\omega(t)), \w)}
       {\partial \w^T}{\omega(t)} (\w - \onevec).
\end{align*}

It will be useful to adopt a specific ``big O'' notation for the remainder of the
proof, by which we mean the following.  If we write $x = O(\sqrt{\alpha})$ for
some quantity $x$, we mean that there exists a constant $C$, available as a
closed-form function of constants defined in \assuref{ij_assu, thetafun_smooth},
such that $x \le C \sqrt{\alpha}$ for all $\alpha \le \frac{\Delta^2}{D\cgh^2}$.
An analogous notation meaning is given to $x = O(\alpha)$.  This ``big O''
notation can be manipulated in the usual ways \citep{bruijn:1981:asymptotic}.

To begin with, by definition of $W_\alpha$, we have $\max_{\w \in W_\alpha}
\meann (\w_n - 1)^2 = \frac{\lfloor \alpha N \rfloor}{N} \le \alpha$, so
$\max_{\w \in W_\alpha} \vnorm{(\w - \onevec) / \sqrt{N}}_2 \le \sqrt{\alpha}$.

Next, observe that \eqref{theta_diff_bound, theta_accuracy_bound} together imply
that
\begin{align*}
\max_{\w \in W_\alpha^*}\vnorm{\fracat{\dee\thetahat(\w)}{\dee\w^T}{\onevec}(\w -
\onevec)}_2
\le \max_{\w \in W_\alpha^*} \vnorm{\thetalin(\w) - \thetahat(\w)}_2 +
\max_{\w \in W_\alpha^*} \vnorm{\thetahat(\w) - \thetahat}_2
= O(\sqrt{\alpha}).
\end{align*}
By \lemref{derivative_smooth} below, we have that
\begin{align*}
\max_{t \in [0,1]}
\max_{\w \in W_\alpha^*}
\vnorm{\left(
    \fracat{\dee \thetahat(\w)}{ \dee \w^T}{\omega(t)} -
    \fracat{\dee \thetahat(\w)}{ \dee \w^T}{\onevec}
    \right) (\w - \onevec) }_2
\le C_b \frac{\vnorm{\w - \onevec}_2}{\sqrt{N}} \sqrt{\alpha}
= O(\alpha).
\end{align*}
Combining the previous two displays gives, by the triangle inequality, that
$\max_{t \in [0,1]} \max_{\w \in W_\alpha^*} \vnorm{\fracat{\dee \thetahat(\w)}{
\dee\w^T}{\omega(t)}(\w - \onevec)}_2 = O(\sqrt{\alpha})$.

Finally, by the Lipschitz property of the partial derivatives in
\assuref{thetafun_smooth}, we have that
\begin{align*}
\max_{t \in [0,1]} \vnorm{
    \fracat{\partial \thetafun(\theta, \omega(t))}
           {\partial \theta}{\thetahat(\omega(t))} -
   \fracat{\partial \thetafun(\theta, \onevec)}
          {\partial \theta}{\thetahat}
}_2
={}& O(\sqrt{\alpha}) \quad \textrm{and}\\
\max_{t \in [0,1]} \sqrt{N} \vnorm{
    \fracat{\partial \thetafun(\thetahat(\omega(t)), \w)}
           {\partial \w}{\omega(t)} -
    \fracat{\partial \thetafun(\thetahat, \w)}
           {\partial \w}{\onevec}
}_2 ={}& O(\sqrt{\alpha}).
\end{align*}
Again, the triangle inequality with the boundedness of the partial
derivatives of $\phi$ at $\w = \onevec$ implies
\begin{align*}
\max_{t \in [0,1]} \vnorm{
    \fracat{\partial \thetafun(\theta, \omega(t))}
           {\partial \theta}{\thetahat(\omega(t))}
}_2 \quad\textrm{and}\quad
\max_{t \in [0,1]} \sqrt{N} \vnorm{
    \fracat{\partial \thetafun(\thetahat(\omega(t)), \w)}
           {\partial \w}{\omega(t)}
}_2
={}& O(\sqrt{\alpha}).
\end{align*}
Combining the above results gives that
\begin{align*}
\max_{t \in [0,1]} \vnorm{\fracat{d \thetafun(\omega(t))}{d t}{t}}_2
= O(\sqrt{\alpha})
\quad\textrm{and}\quad
\max_{t \in [0,1]} \vnorm{
    \fracat{\dee \thetafun(\omega(t))}{\dee t}{t} -
    \fracat{\dee \thetafun(\omega(t))}{\dee t}{1}
}_2  = O(\alpha),
\end{align*}
from which the desired conclusion follows by \eqref{thetafun_integral}.
\qed


%% file: appendix_bound_tight.tex
Recall that \thmref{thetafun_accuracy} provides upper bounds on
two quantities:
\begin{align*}
    \textrm{Error} :={}&
    \abs{\thetafunlin(\w) - \thetafun(\thetahat(\w), \w)} \le C_1 \alpha
&
    \textrm{Diff} :={}&
        \abs{\thetafun(\thetahat(\w), \w) - \thetafunhat}  \le C_2 \sqrt{\alpha}.
\end{align*}
To show that \thmref{thetafun_accuracy} is tight, we will construct a sequence
of Z-estimators which satisfy \assuref{ij_assu} with a particular set of
constants for all $N$, and which satisfies
\begin{align*}
\frac{\textrm{Error}}{\alpha} ={}&
\frac{\abs{\thetafunlin(\w) - \thetafun(\thetahat(\w), \w)}}{\alpha}
\rightarrow{} 1 
&
\frac{\textrm{Difference}}{\sqrt{\alpha}} ={}&
\frac{\abs{\thetafun(\thetahat(\w), \w) - \thetafunhat}}{\sqrt{\alpha}}
\rightarrow{} 1
\end{align*}
as $\alpha \rightarrow 0$ (which requires $N \rightarrow \infty$ since we will
require $\alpha > 1/N$ to leave out at least one datapoint).  As $\alpha$
approaches zero, this Z estimator comes arbitrarily close to the upper bounds of
\thmref{thetafun_accuracy}, showing that \thmref{thetafun_accuracy} is tight.

For a scalar $\theta$, we consider a Z-estimator of the form
\begin{align}\eqlabel{tight_z_estimator}
G(\theta, \d_n)  = \theta b_n - a_n,
\end{align}
where $\d_n = (a_n, b_n)$.\footnote{Such an estimating equation might arise from
minimizing the squared error loss $\sumn \frac{1}{2} (x_n \theta - y_n)$, where
we take $b_n = x_n^2$ and $a_n = x_n y_n$. For any $b_n > 0$ and any $a_n$, we
can define a corresponding $x_n$ and $y_n$ that would give rise to
\eqref{tight_z_estimator}.} Let $a = (a_1, \ldots, a_N)$ and $b=(b_1, \ldots,
b_N)$ denote the corresponding $N$-vectors. We will shortly choose the values of
$a$ and $b$ carefully, but we will immediately assume that $\sumn a_n =
\onevec^T a = 0$. We will take $\theta$ itself to be our quantity of interest,
i.e., $\thetafun(\theta)  = \theta$.  

It follows by direct computation that
\begin{align*}
\thetafun(\w) = \thetahat(\w) ={}& \frac{a^T \w}{b^T \w}
&
\fracat{\partial \thetafun(\w)}{\partial \w^T}{\w} ={}&
    - \frac{\thetahat(\w) b^T - a^T}{ b^T \w}\\
\thetafunhat = \thetafun(\onevec) ={}& 0
&
\thetafunlin(\w) ={}& \frac{a^T \w}{ b^T \onevec}
\end{align*}

Plugging in, we see that
\begin{align*}
\textrm{Error} ={}&
\abs{a^T \w}\abs{\frac{1}{ b^T \onevec} - \frac{1}{b^T \w}}
&
\textrm{Diff} ={}&
\abs{\frac{a^T \w}{b^T \w}}.
\end{align*}
We will take $\w_\alpha$ to be a weight vector that is zero in its first $\alpha
N$ entries and one otherwise, for $\alpha \propto 1/N$ and $\alpha \in (0,
1/2)$.

Without additional constraints, the error and difference can take a wide range
of values for a particular $\w$.  For example, if $a$ is identically zero, then
both are zero.  Alternatively, if $b^T w = b^T \onevec$, then the error is zero,
even if the difference is not.  More pathologically, if $b^T \onevec \ll b^T
\w$, then the error can be arbitrarily large relative to the difference.  
In practical terms, this pathological setting would correspond to a nearly
singular objective Hessian, and very large $\cop$ in \assuref{ij_assu}.

However, this pathological case does not contradict \thmref{thetafun_accuracy}.
\Thmref{thetafun_accuracy} states that, for a particular set of constants given
by \assuref{ij_assu}, as $\alpha \rightarrow 0$, the bounds of
\thmref{thetafun_accuracy} apply.  Therefore, to investigate the tightness of
the bounds in \thmref{thetafun_accuracy}, we will choose $a$ and $b$ in such a
way that, as $N$ grows, \assuref{ij_assu} is satisfied for a particular set of
reasonable constants for all $N$.  It will suffice to choose $a$ and $b$
satisfying
\begin{align}
    N^{-1} a^T \onevec ={}& 0 &
    N^{-1} a^T a ={}& 1 \nonumber\\
    N^{-1} b^T \onevec ={}& 1   &
    N^{-1} (b - \onevec)^T (b - \onevec) ={}& 1.
    \eqlabel{z_constraints}
\end{align}
Equivalently, we require that the $a_n$ have sample mean equal to $0$, the $b_n$
have sample mean equal to $1$, and both have sample variance equal to $1$.

Let $\w_\alpha$ denote a weight vector which is all ones except for zeros in the
first $\alpha N$ entries, for $\alpha \propto 1/N$ and $\alpha > 0$. Within
these constraints, we will attempt to choose $a$ and $b$ so that the error and
difference are both as large as possible for a given $\w_\alpha$. For the
remainder of this section, suprema and infima taken over $a$ and $b$ will
implicitly be taken over $a$ and $b$ satisfying \eqref{z_constraints}.

First, we choose $a$. Both the error and difference are made large by making
$a^T \w_\alpha$ large for a given $\w_\alpha$, subject to $a^T \onevec = 0$ and
$\frac{1}{N} a^T a = 1$. We solve this constrained optimization problem in
\pointref{shape} of \secref{amip_decomposition}, showing that 

\def\astar{{a^{*}}}
\def\bstar{{b^{*}}}
\begin{align*}
\sup_{a} a^T \w_\alpha = \astar^T \w_\alpha = N \sqrt{\alpha ( 1 - \alpha)},
\end{align*}
where $\astar$ is constant and negative in entries where $\w_\alpha$ is zero,
and constant and positive in entries where $\w_\alpha$ is one.

\def\bbar{\bar{b}}
\def\wdiff{\Delta_\alpha}

Next, to choose $b$, we form a series expansion of $(b^T \w_\alpha)^{-1}$.
Define $\bbar := b - \onevec$ (so that $\onevec^T \bbar = 0$ and $N^{-1}
\bbar^T \bbar = 1$), and define $\wdiff := \w_\alpha - \onevec$.  
For sufficiently small $\alpha$, we can form the following expansion:
\begin{align*}
\frac{1}{b^T \w_\alpha} ={}&
\frac{1}{N} \frac{1}{\frac{1}{N} b^T (\wdiff + \onevec)} =
\frac{1}{N} \frac{1}{\frac{1}{N} b^T \wdiff + 1} =
\frac{1}{N} \left(1 - \frac{1}{N} b^T \wdiff + \left(\frac{1}{N} b^T \wdiff\right)^2 - 
    \ldots \right).
\end{align*}
The expansion is justified for sufficiently small $\alpha$ by \pointref{shape}
of \secref{amip_decomposition}, since, given these constraints on $\bbar$,
\begin{align*}
\sup_{b} \frac{1}{N} \abs{b^T \wdiff} ={}&
\sup_{\bbar} \frac{1}{N} \abs{(\bbar + \onevec )^T \w_\alpha  - N} ={}
\sup_{\bbar} \frac{1}{N} \abs{\bbar^T \w_\alpha  -\alpha N} \le {}
\abs{\sqrt{\alpha(1 - \alpha)}} + \alpha.
\end{align*}
The preceding inequality also implies that, for sufficiently small $\alpha$, the
leading term will dominate, and this leading term will be made as large as
possible by making $\frac{1}{N} b^T \wdiff$ as negative as possible.  By the
same reasoning as the preceding display,
\begin{align*}
\inf_{b} \frac{1}{N} b^T \wdiff = - \sqrt{\alpha(1 - \alpha)} - \alpha,
\end{align*}
where the supremum is achieved at a value we call $\bstar$.
Applying the same expansion, we analogously get
\begin{align*}
    \frac{1}{b^T \w} -     \frac{1}{ b^T \onevec} ={}&
    \frac{1}{N} \left(- \frac{1}{N} b^T \wdiff + \left(\frac{1}{N} b^T \wdiff\right)^2 - 
    \ldots \right).
\end{align*}
Consequently, the leading order term of $\abs{\frac{1}{ b^T \onevec} -
\frac{1}{b^T \w}}$ is also maximized at $\bstar$.
Combining, we see that, as  $\alpha \rightarrow 0$,
\begin{align*}
\frac{N}{\bstar^T \w_\alpha} \rightarrow{}& 1
&
\frac{N \abs{\frac{1}{ \bstar^T \onevec} -
             \frac{1}{\bstar^T \w_\alpha}}}{\sqrt{\alpha}} \rightarrow{}& 1.
\end{align*}

Evaluating both the difference and error at $\astar$ and $\bstar$, we get
our desired result:
\begin{align*}
\frac{\textrm{Error}}{\alpha} ={}&
\frac{\abs{\astar^T \w_\alpha}\abs{
    \frac{1}{ \bstar^T \onevec} - \frac{1}{\bstar^T \w_\alpha}}}{\alpha}
    \rightarrow 1
&
\frac{\textrm{Difference}}{\sqrt{\alpha}} ={}&
\frac{\abs{\frac{\astar^T \w}{\bstar^T \w_\alpha}}}{\sqrt{\alpha}}
\rightarrow 1.
\end{align*}